%% file: springer.tex
\documentclass[runningheads,a4paper]{llncs}

\usepackage{amssymb}
\usepackage{graphicx}

\usepackage{url}
\usepackage[caption=false]{subfig}

\urldef{\mailsa}\path|{alfred.hofmann, ursula.barth, ingrid.haas, frank.holzwarth,|
\urldef{\mailsb}\path|anna.kramer, leonie.kunz, christine.reiss, nicole.sator,|
\urldef{\mailsc}\path|erika.siebert-cole, peter.strasser, lncs}@springer.com|    
\newcommand{\keywords}[1]{\par\addvspace\baselineskip
\noindent\keywordname\enspace\ignorespaces#1}

\begin{document}

\mainmatter  

\title{Who is Dating Whom: Characterizing User Behaviors of a Large Online Dating Site}


%
%
\author{Peng Xia\inst{1} \and Kun Tu\inst{2} \and Bruno Ribeiro\inst{2} \and Hua Jiang\inst{3} \and Xiaodong Wang\inst{3} \and \\
Cindy Chen\inst{1} \and Benyuan Liu\inst{1} \and Don Towsley\inst{2}}

%

\institute{Department of Computer Science, University of Massachusetts Lowell, Massachusetts, USA\\ \and
Department of Computer Science, University of Massachusetts Amherst, Massachusetts, USA\\ \and
Product Division, Baihe.com, Beijing, China
}

%
%

\toctitle{Lecture Notes in Computer Science}
\tocauthor{Authors' Instructions}
\maketitle

\begin{abstract}
Online dating sites have become popular platforms for people to look for potential romantic
partners. It is important to understand users' dating preferences in order to make better
recommendations on potential dates. The message sending and replying actions of a user are 
strong indicators for what he/she is looking for in a potential date and reflect the user's 
actual dating preferences. We study how users' online dating behaviors correlate with various 
user attributes using a large real-world dateset from a major online dating site in China. Many
of our results on user messaging behavior align with notions in social and evolutionary 
psychology: males tend to look for younger females while females put more emphasis on the 
socioeconomic status (e.g., income, education level) of a potential date. In addition, we observe
that the geographic distance between two users and the photo count of users play an important role 
in their dating behaviors. Our results show that it is important to differentiate between users' 
true preferences and random selection. Some user behaviors in choosing attributes in a potential 
date may largely be a result of random selection. We also find that both males and females are more
likely to reply to users whose attributes come closest to the stated preferences of the receivers, 
and there is significant discrepancy between a user's stated dating preference and his/her actual 
online dating behavior. These results can provide valuable guidelines to the design of a recommendation 
engine for potential dates.
\end{abstract}

\input{introduction}
\input{relatedWork2}

\input{dataset}
\input{replyProb2}
\input{conclusion}

\label{references}

\end{document}

%% file: introduction.tex
\section{Introduction}
\vspace*{0.1375\baselineskip}
Computer-based matchmaking was pioneered by Operation Match at Harvard University and 
Contact at MIT in mid-1960s \cite{Slater13Love}. Based on the responses to a personality 
questionnaire, a computer program tried to match a user with compatible dates. Three decades later,
starting in the mid-1990s, with the increasing ubiquity of the Internet connectivity and wide-spread
use of the World Wide Web, online dating sites have emerged as popular platforms for people 
to look for potential romantic partners. 

The rise of online dating has fundamentally altered the dating landscape and profoundly impacted 
people's dating life. It offers an unprecedented level of access to potential romantic partners that 
is otherwise not available through traditional means. According to a recent 
survey\footnote{http://statisticbrain.com/online-dating-statistics}, 
40 million single people (out of 54 million) in the US have signed up with various online dating sites such 
as Match.com, eHarmony, etc, and around 20\% of currently committed 
romantic relationships began online, which is more than through any means other than meeting 
through friends. 


An online dating site allows a user to create a profile that typically includes the user's photos,
basic demographic information, behavior and interests (e.g., smoking, drinking, hobbies), 
self-description, and desired characteristics of an ideal partner. Some sites require a user to 
complete a personality questionnaire for evaluating the person's personality type and using it in 
the matching process. After creating a profile, a user can  search for other 
people's profiles based on a variety of user attributes, browse other user profiles, and exchange 
messages with them. Many sites provide suggestions on compatible partners based on proprietary 
matching algorithms.

There is often considerable discrepancy, or dissonance (a concept in social psychology), between 
a user's stated preference and his or her actual dating behavior \cite{Eastwick08Sex}. Therefore, it 
is important to understand users' true dating preferences in order to make better dating recommendations. 
The message send and reply actions of a user are strong indicators for what he/she is looking
for in a potential partner and reflect the user's actual dating preferences. 

In this paper we study how user online dating behavior correlates with various user attributes using a 
real-world dateset obtained through a collaboration with baihe.com, one of the largest online dating sites 
in China with a total number of 60 million registered users. In particular, we address the following research 
questions: 
\begin{itemize}
\item {\it Temporal behaviors}: How often does a user send and receive messages and how 
does this change over time? How long does it take a recipient to reply to a message he/she received? 

\item {\it Send behaviors}: What is the relationship between the attributes of initiators and 
recipients of the initial messages?  How does user messaging behavior differ from random 
selection? How do users' actual online dating behaviors deviate from their stated preferences?

\item  {\it Reply behaviors}:  How does the reply probability of a message correlate with various 
attributes of the sender and receiver? How does the reply probability depend on the extent to which
the sender's attributes match the receiver's stated preferences?

\end{itemize}


\noindent {\bf Main findings:} 
Our study provides a firsthand account of the user online dating behaviors based on a large
dataset obtained from a large online dating site (baihe.com) in China, a country with a very
large population and unique culture. On average, a male sends out more messages but receives
fewer messages than a female. A female is more likely to be contacted but less likely 
to reply to a message than a male. The number of messages that a user sends out and 
receives per week quickly decreases with time. Most messages are replied to within a
short time frame with a median delay of around 9 hours. 

Many of our results on user messaging behavior align with notions in social 
and evolutionary psychology \cite{Buss89Sex} \cite{Eagly99Origin} \cite{Luo05Assortative}. 
Males tend to look for younger females while females place more emphasis on 
socioeconomic status such as the income and education level of a potential date. 
As a male gets older, he searches for relatively younger and younger women.  A 
female in her 20's is more likely to look for older males, but as a female gets older, 
she becomes more open towards younger males.

In addition to the above findings, we observe that geographic distance between two 
users plays an important role in online dating considerations: 46.5\% of the initial 
messages occurred between users in the same city, and for messages that cross the city boundaries, 
the volume quickly decreases as users live farther apart. Females are more likely 
than males to send and reply to messages between distant big cities. Profile 
photos affect male and female's messaging behaviors differently. Females with a 
larger number of photos are more likely to invite messages and secure replies from males, 
but the photo count of males does not have as significant effect in attracting contacts and 
replies. 

Our results also show that it is important to differentiate between users' true preferences 
and random selection. Some user behaviors in choosing attributes in a potential date may 
be a result of random selection. For example, while it appears 
that a male tends to look for females shorter than he is and a female tends to look for males 
taller than she is, the message send and reply behaviors of both genders closely approximate 
those resulting from random selection, showing that these behaviors may result from random 
selection rather than users' true preferences. 

Our results also indicate a significant discrepancy between a user's stated dating
preference and his/her actual online dating behavior. A fairly large fraction of messages 
are sent to or replied to users whose attributes do not match the sender or receiver's stated 
preferences. Females tend to be more flexible than males in deviating from their stated preferences 
when sending and replying to messages. For both males and females, out of the population of users 
that send messages, replies are more likely to go to users whose attributes come closest to the 
stated preferences of the receivers.

In summary, our results reveal how user message send and reply behaviors correlate with 
various user attributes, how these behaviors differ from random selection, and how users' actual 
online dating behavior deviates from their stated preferences. These results on users' dating 
preferences can provide valuable guidelines to the design of recommendation engine for potential 
dates. 

The rest of the paper is structured as follows. Section \ref{sec:relatedWork} presents an
overview of previous studies on the data analysis of online dating sites. Section 
\ref{sec:dataset} describes the  dataset that we obtained from a major online dating site in China.  
Section \ref{sec:temporal} describes the temporal characteristics of users' online dating behavior. 
Users' message send and reply behaviors are studied in Section \ref{sec:behavior}. We discuss
our main results in Section \ref{sec:discussions}. Finally, we conclude the paper in Section \ref{sec:conclusion}.



%% file: relatedWork2.tex
\section{Related Work} \label{sec:relatedWork}
Hitsch et al.~\cite{Hitsch10What} shows that in online dating there is no evidence for user strategic
behavior shading their true preference. Both male and female users have a strong preference for 
similarity along many (but not all) attributes. U.S.\ users display strong same-race correlations.
There are gender differences in mate preferences; in particular, women have a stronger preference
than men for income over physical attributes. In their follow-up work~\cite{Hitsch10Matching} they 
show that stable matches obtained through the Gale-Shapley algorithm are similar to the actual matches 
achieved by the dating site, which are also approximately efficient. 

Fiore et al.~\cite{Fiore10Who} analyzes messaging behavior and find them consistent with predictions 
from evolutionary psychology, women state more restrictive preferences than men and contact and reply 
to others more selectively. Lin et al.~\cite{Lin13Mate} studied how race, gender, and education jointly 
shape interaction among heterosexual Internet daters. They find that racial homophily dominates mate 
searching behavior for both men and women. This is not the case of Chinese online daters where the 
overwhelming majority of users are of the same race. Finkel et al. \cite{Finkel12Online} states that 
online dating has fundamentally altered the dating landscape by offering an unprecedented level of 
access to potential partners and allowing users to communicate before deciding whether to meet them 
face-to-face. On the other hand, the authors also argue that there is no strong evidence that
matching algorithms promote better romantic outcomes than conventional offline dating. Part of 
the problem is that the main principles underlying these algorithms (typically similarity but also
complementarity) are much less important to relationship well-being than online sites are willing to assume. 
Interesting on-the-fly statistics of OKcupid users is found at the OkTrends blog~\cite{OkTrends}.

%% file: dataset.tex
\section{Dataset Description} \label{sec:dataset}

We report on a dataset taken from baihe.com, a major online dating site in China. It includes the profile information of 200,000 users uniformly sampled from users registered in November 2011. For each user, we have his/her message sending and receiving traces (who contacted whom at what time) in the online dating site and the profile information of the users that he or she has communicated with from the date that the account was created until the end of January 2012.

A user's profile provides a variety of information including user's gender, age, current location (city and province), home town location, height, 
weight, body type, blood type, occupation, income range, education level, religion, astrological sign, marriage and children status, number of photos 
uploaded, home ownership, car ownership, interests, smoking and drinking behavior, self introduction essay, among others. Each user also provides his/her preferences for potential romantic partners in terms of age, location, height, education level, income range, marriage and children status, etc.









\begin{figure}[htb]
\centering
\includegraphics[width=4.0in]{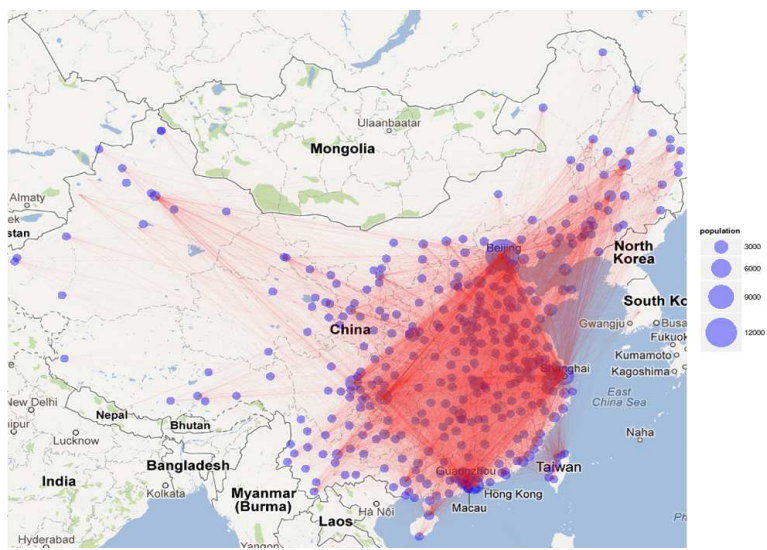}
    \caption{Inter-city communications of the online dating site within China.}
   \label{fig:baiheMessage}
\end{figure}

Of the 200,000 sampled users, 139,482 are males and 60,518 are females, constituting 69.7\% and 30.3\% of the total number of sampled users respectively. 
The dataset includes people from 34 countries and all of the provinces and municipalities (cities directly under the jurisdiction of the central government including Beijing, Shanghai, Tianjin, Chongqing), and special administrative region (Hong Kong, Macau) in China. Figure \ref{fig:baiheMessage} illustrates the user geographical locations (at city level) within China and the inter-city communications between users. Intra- and inter-city messages constitute 46.5\% and 53.5\% of the total message volume in our data, respectively.

\begin{figure}
\subfloat[\label{fig:ageDistribution}]
  {\includegraphics[width=.5\linewidth]{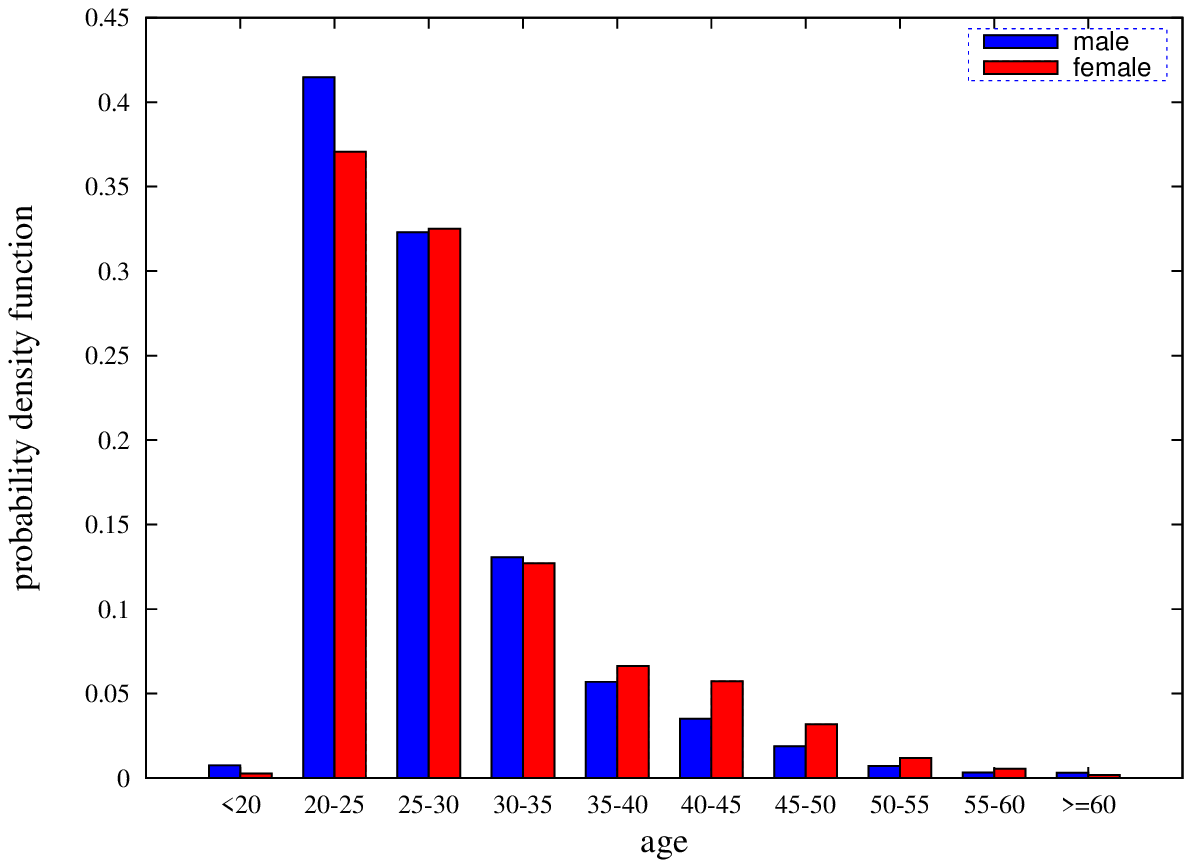}}\hfill
\subfloat[\label{fig:incomeCDF}]
  {\includegraphics[width=.5\linewidth]{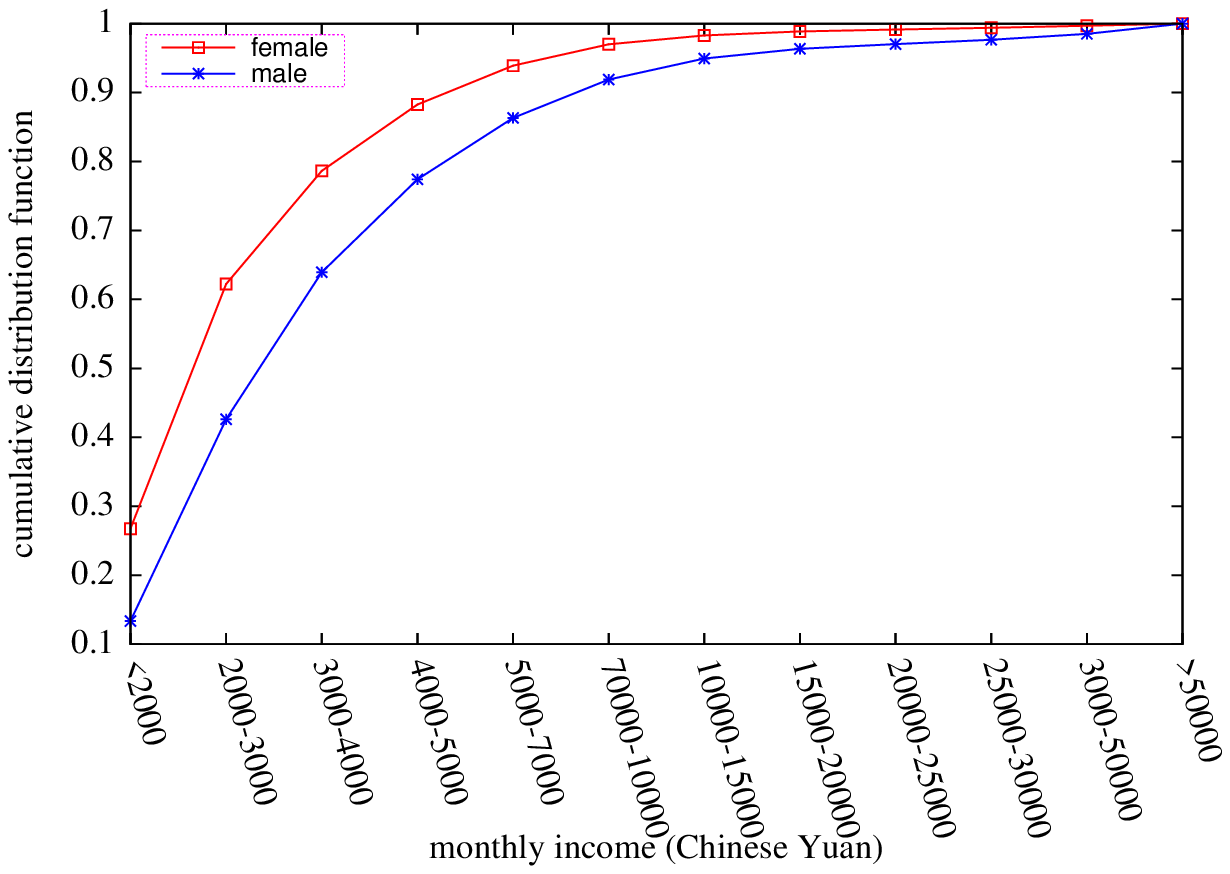}}\hfill
\subfloat[\label{fig:educationDistribution}]
  {\includegraphics[width=.5\linewidth]{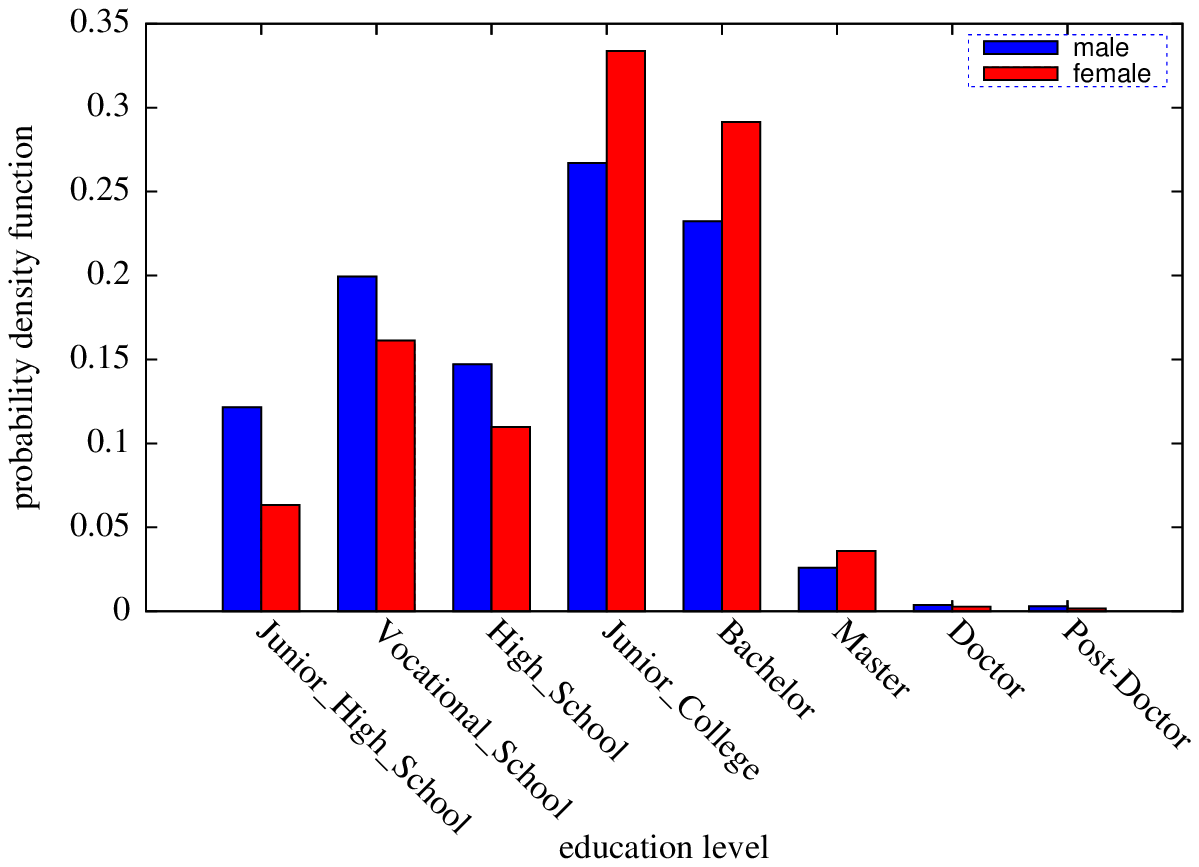}}\hfill
\subfloat[\label{fig:marriageStatus}]
  {\includegraphics[width=.5\linewidth]{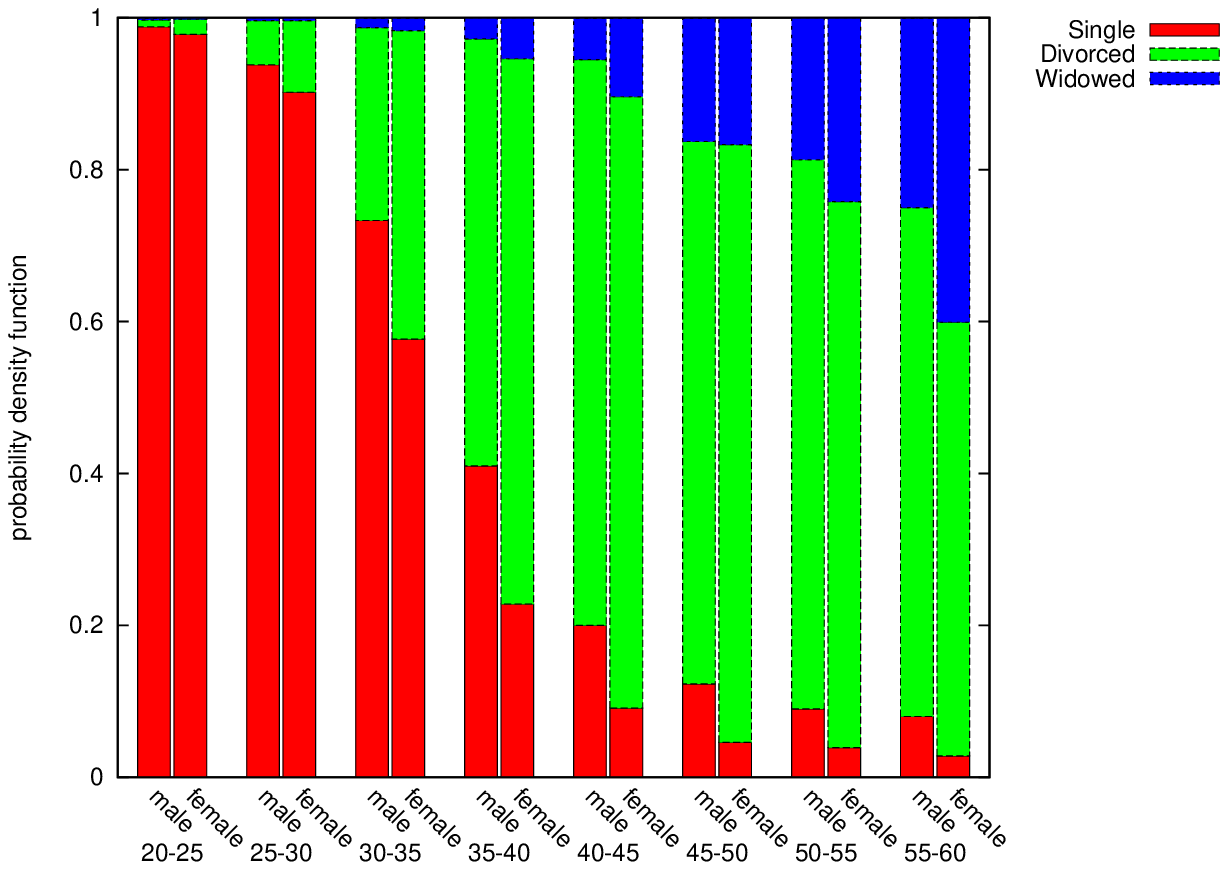}}
\caption{(a) Age distribution of the male and female users. (b) Cumulative distribution function of users' reported monthly income. (c) Education level distribution of the male and female users. (d) Marriage status distribution of the male and female users.}
\label{fig:sample}
\end{figure}

To give a sense of the main user demographic attributes, we plot distributions of user reported age, 
education level, monthly income range and marriage status in Figures \ref{fig:sample}\subref{fig:ageDistribution}, 
\ref{fig:sample}\subref{fig:incomeCDF}, \ref{fig:sample}\subref{fig:educationDistribution} and \ref{fig:sample}\subref{fig:marriageStatus}, respectively.


The youngest user is 19 years old and the largest fraction of users are in their early 20s. While there is a larger fraction of male users than female users below age 25, the fraction of female users starts to match that of male users for age range 25-35, and exceeds that of male users after age 35. The median ages of male and female users are 25 and 26, respectively.




The fraction of female users is larger than that of male users for low income ranges (less than 3,000 Chinese Yuan per month). For higher income ranges, the 
trend becomes opposite. In general, males have larger incomes than females in our dataset. The median income ranges of male and female users are 3,000-4,000 and 
2,000-3,000 Chinese Yuan, respectively. 

With respect to users' education level, females stated education levels tend to be higher than males. About 66.5\% of females state that they have at least a community college degree in contrast with only 53.2\% of the males. The fraction of users with stated doctoral and post-doctoral degrees is 0.61\%.


\vspace*{0.1375\baselineskip}
As shown in Figure \ref{fig:sample}\subref{fig:marriageStatus}, the majority of users in their 
early 20s are singles. As the user age increases, the ratio of single users 
decreases while the ratio of widowed users increases. The ratio of divorced 
users first increases with the user ages until mid-40s and then starts to 
decrease. In general, the ratios of widowed and divorced female users are 
larger than those of male users.


\vspace*{0.1375\baselineskip}
Unlike online dating behaviors in US where race plays an important role when it comes to finding potential romantic partners \cite{OkTrends} \cite{Lin13Mate}, most of the users (98.9\%) in our dataset are Han (ethnic majority in China), and all other ethnic groups comprise 1.1\% of the users. Moreover, the majority of the users (97.0\%) claim to be non-religious. Those claiming a religion (Buddhism, Taoism, Catholic, Islamism, etc) constitute only 3.0\% of users. Note that the race and religion compositions in our dataset are significantly different from those of online dating sites in the US where there is more diversity \cite{Lin13Mate} \cite{Hitsch10What}.

\begin{figure}
\subfloat[\label{fig:weeklySender}]
  {\includegraphics[width=.5\linewidth]{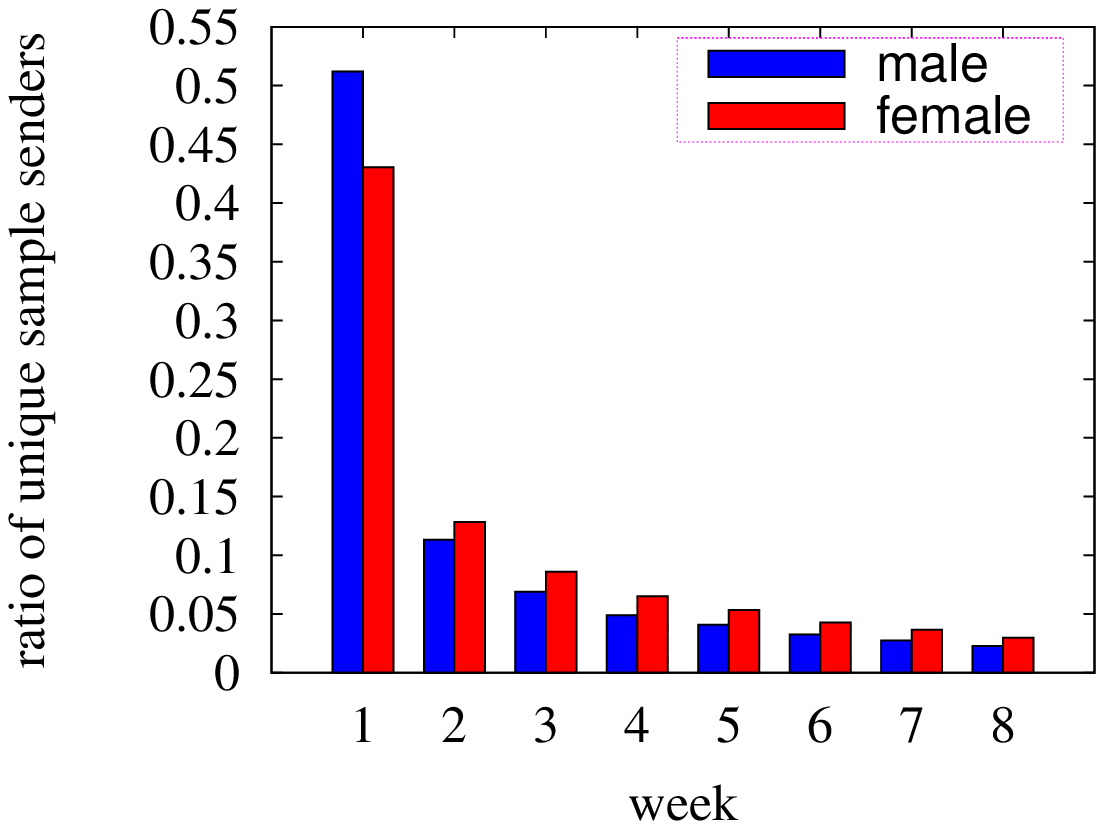}}\hfill
\subfloat[\label{fig:weeklySentMessages}]
  {\includegraphics[width=.5\linewidth]{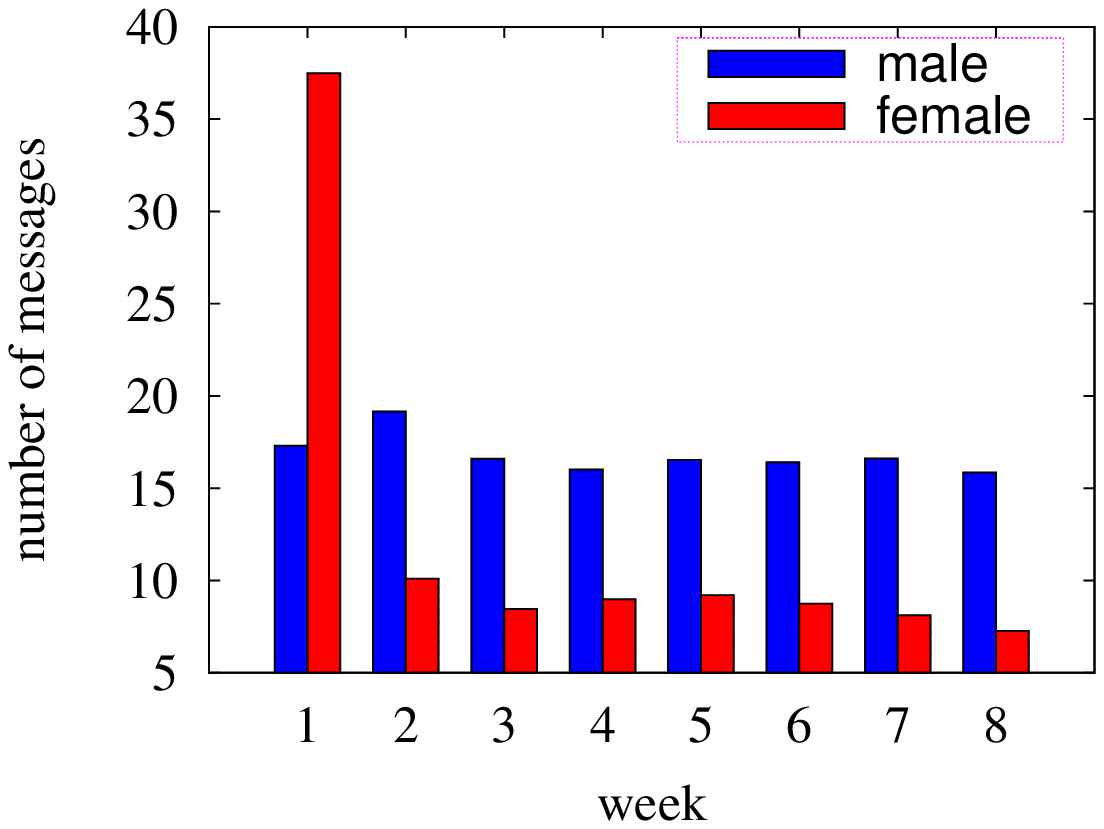}}
\caption{(a) Fraction of users who sent out at least one message during a week. (b) Average number of messages a user sent out each week given that a user sends at least one message.}
\label{fig:weeklySend}
\end{figure}

For each user in our sample, we have the time stamps of the messages as well as the profile information of users that this user has communicated with. In this paper we focus on the initial messages exchanged between users. Subsequent messages between the same pair of users do not represent a new sender-receiver pair and cannot be used as the only indicator for continuing relationship as users may choose to go off-line from the site and communicate via other channels (e.g., email, phone, or meet in person).

\section{Temporal Behaviors} \label{sec:temporal}

We are interested in how a user's online dating activity level changes over time after he or she registers an account on the online dating site. Since we only have eight full weeks' worth of online dating data for users who joined in November 30, 2011, we only consider the activities of each user during the first eight weeks of his/her membership. The following analysis is based on the activities of the 200,000  users in the dataset described in Section \ref{sec:dataset}.

{\bf Messages sent from sample users}: During the eight-week period, 2,089,029 initial messages were sent by 76,654 males (55.0\% of the males in the dataset) to 508,118 unique females, which in turn generated 156,774 replies (a reply rate of 7.5\%). During the same time period, 1,217,672 initial messages were sent by 29,535 females (48.8\% of the females in the dataset) to 440,714 unique male users, which in turn generated 112,696 replies (a reply rate of 9.3\%). 

{\bf Messages sent to sample users}: During the same time period, 328,645 initial messages were sent by 94,179  females  to 44,509  males, which in turn generated 58,946 replies (a reply rate of 17.9\%). 1,586,059 initial messages were sent by 288,602 males to 45,623 females, which in turn generated 150,917 replies (a reply rate of 9.5\%). Note that males are more likely to initiate contact than females while messages from females are more likely to generate replies than those from males.








The fraction of users from the dataset that sent out at least one message and the average number of messages sent by each user are shown in Figures \ref{fig:weeklySend}\subref{fig:weeklySender} and  \ref{fig:weeklySend}\subref{fig:weeklySentMessages}, respectively. We observe that while a considerable fraction of users (51.2 \% of males and 43.0\% of females) sent out at least one message during the first week of their memberships, the fraction decreases sharply in the second week (down to  11.3\% for males and 12.8\% for females) and further decreases in subsequent weeks. Except for the first week, females are slightly more likely to send out a message than males on average. The average number of messages a male sends out each week given that he sends at least one message lies between 15 and 20 messages per week.  While the average number of messages a female  sends given that she sends at least one message is more than twice that of a male in the first week, it decreases sharply in the second week and remains relatively stable at a much lower level than that of a male  over the next seven weeks.

\begin{figure}
\subfloat[\label{fig:MessageSentCCDF}]
 {\includegraphics[width=.5\linewidth]{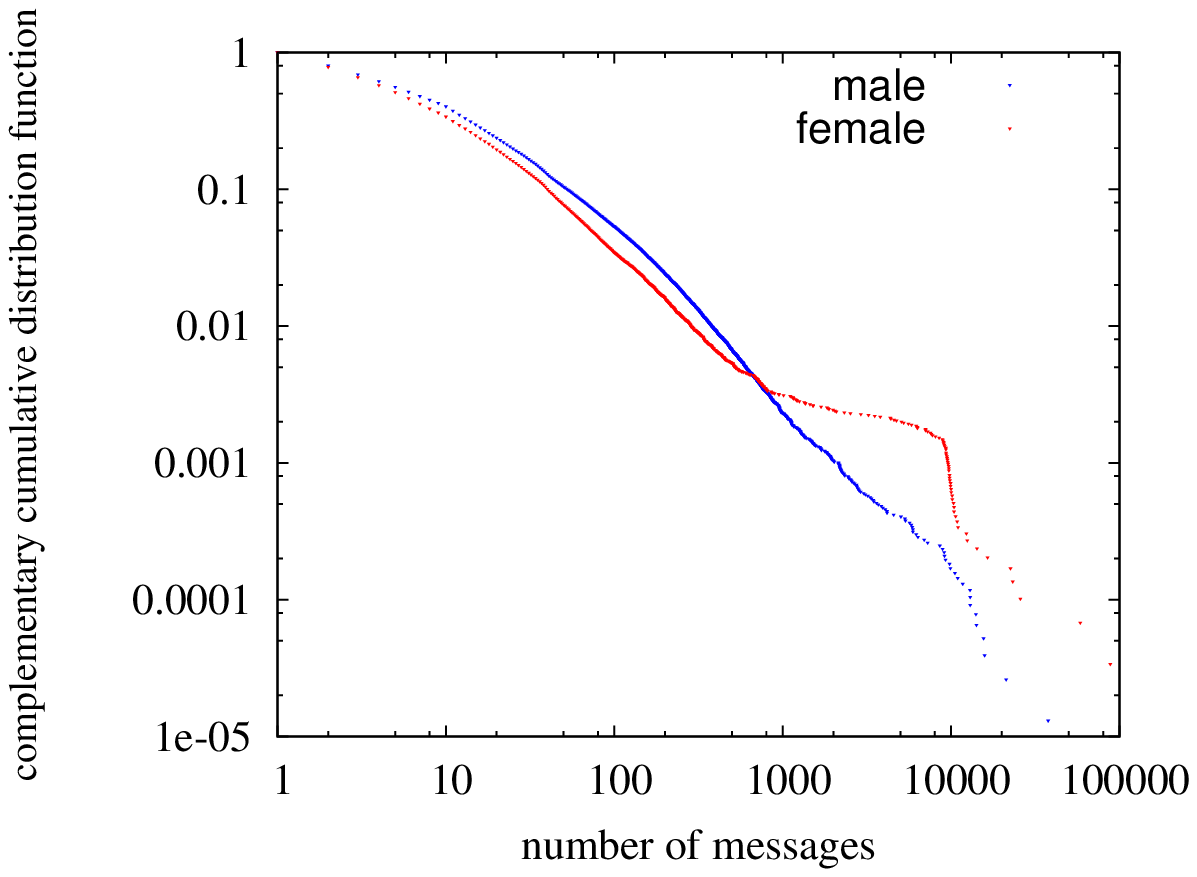}}\hfill
\subfloat[\label{fig:MessageReceivedCCDF}]
  {\includegraphics[width=.5\linewidth]{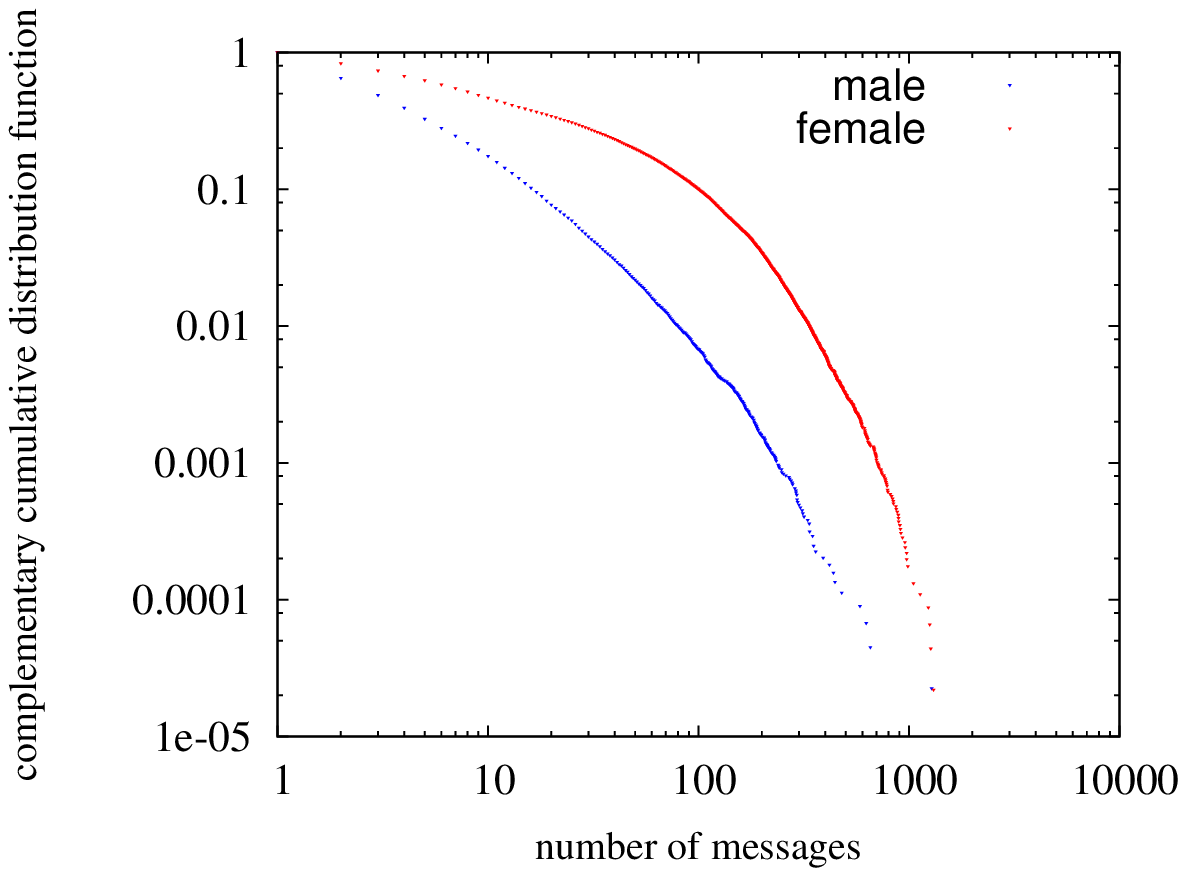}}
\caption{(a) CCDF of the number of messages a user sent out during the first eight weeks of his/her membership. (b) CCDF of the number of messages a user received during the first eight weeks of his/her membership.}
\label{fig:messageCCDF}
\end{figure}

%
%
%

For both males and females, we obtain the distribution of the number of messages sent by each user per week given that a user sends at least one message during the week, and plot its complementary cumulative density function (CCDF) in Figure \ref{fig:messageCCDF}\subref{fig:MessageSentCCDF}. We observe that the distributions exhibit  heavy tails. Most users only sent out a small number of messages: 94.6\% of males and 96.5\% of females  sent out less than 100 messages during the first eight weeks of their membership. On the other hand, there are small fractions of users that sent out a large number of messages. According to the online dating site, most of these highly active users are likely to be fake identities created by spammers and their accounts have been quickly removed from the site.

\begin{figure}
\subfloat[\label{fig:weeklyReceiver}]
  {\includegraphics[width=.5\linewidth]{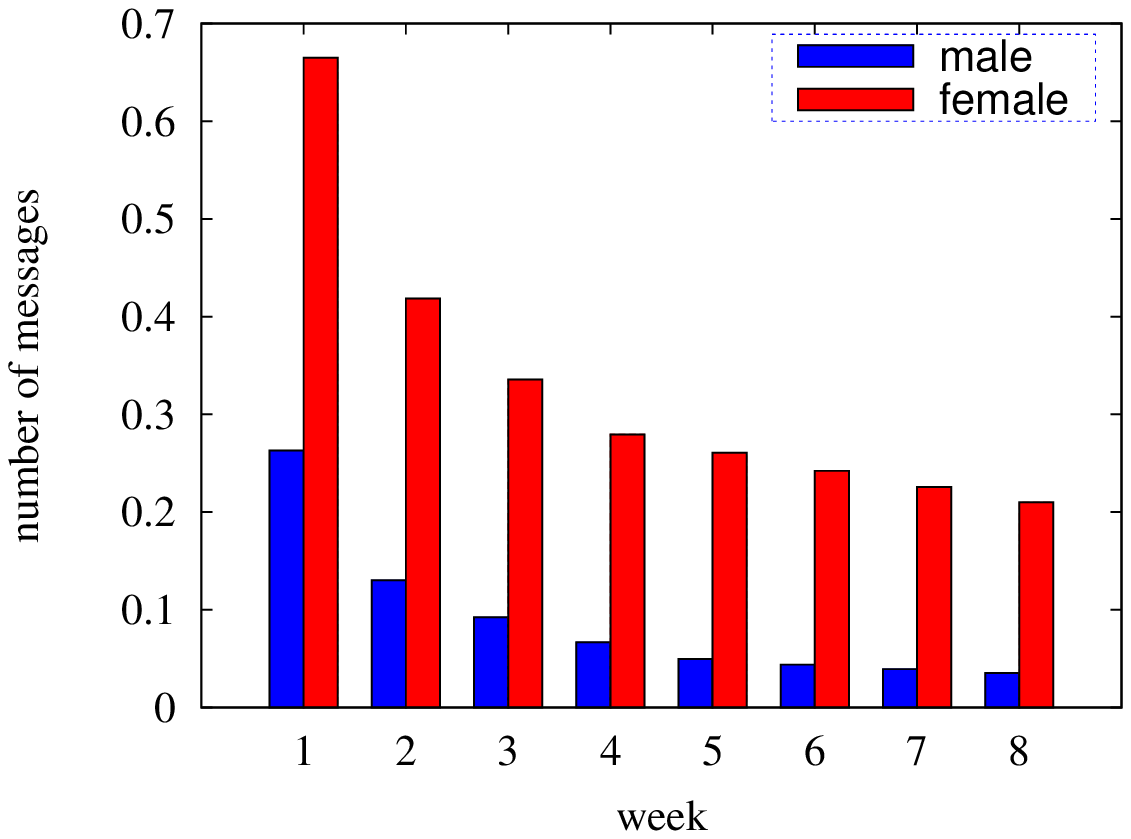}}\hfill
\subfloat[\label{fig:weeklyREeceivedMessages}]
  {\includegraphics[width=.5\linewidth]{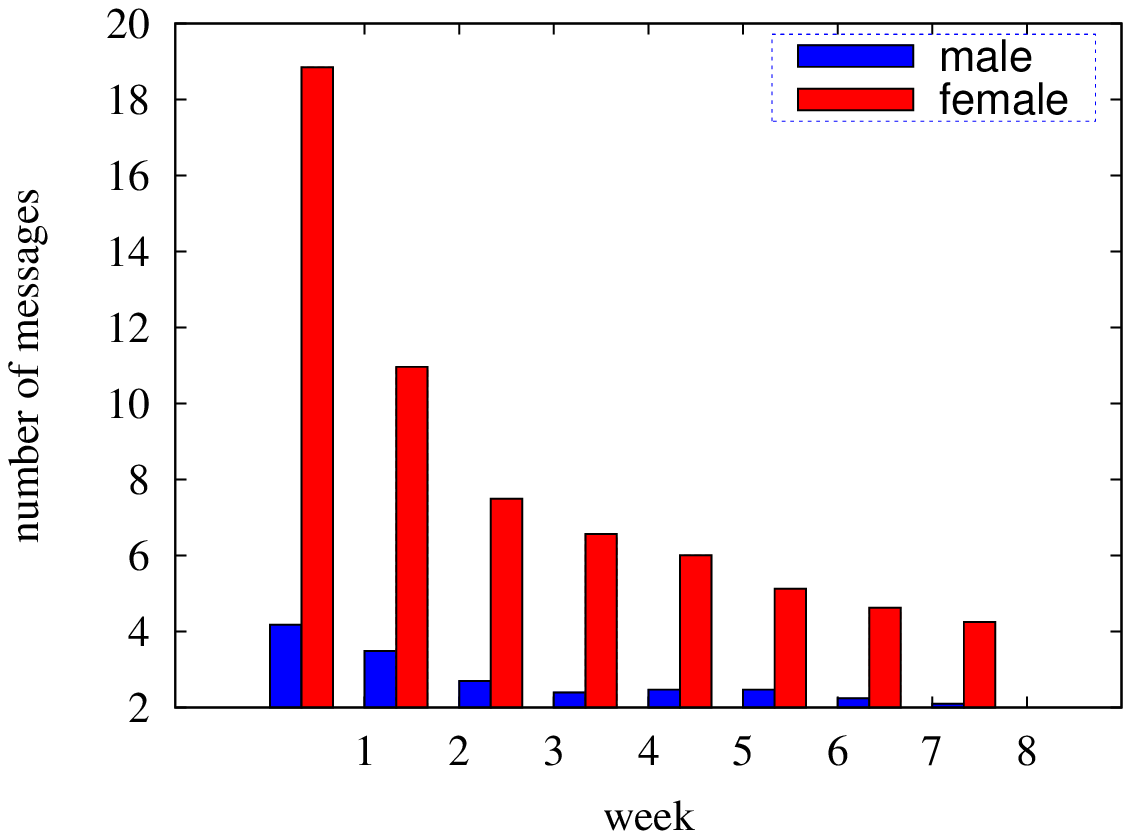}}
\caption{(a) Fraction of users who received at least one message during a week. (b) Average number of messages a user received each week given that a user received at least one message.}
\label{fig:weeklyReceive}
\end{figure}

The fraction of users from the dataset that received at least one message and the average number of messages received by each user during the first eight weeks of his/her membership are shown in Figures \ref{fig:weeklyReceive}\subref{fig:weeklyReceiver} and \ref{fig:weeklyReceive}\subref{fig:weeklyREeceivedMessages}, respectively. We observe that the fractions of both males and females that receive at least one message each week gradually decreases over time, and that females are much more likely to receive  messages than males. Also, for each week, the average number of messages a user received generally decreases over time for both genders, and the number of messages received by a female each week is much larger than that for a male.

%
%

For those users that received at least one message during the first eight weeks of their membership, we show  the complementary cumulative distribution function (CCDF) of the number of messages received by each user for both males and females in Figure \ref{fig:messageCCDF}\subref{fig:MessageReceivedCCDF}. We observe that the distributions for both male and female users exhibit a log-normal-like behavior, and that females tend to receive more messages than male users.


To investigate how long it takes a user to reply after receiving a message, we define the reply delay of a message to be the time elapsed from when the message is sent until the corresponding reply is generated when there is a reply. The reply delay may have certain psychological implications to some people and hence affect the progress of the communication. Thus it is an important metric to study.

\begin{figure}[htb]
\centering
\includegraphics[width=2.2in]{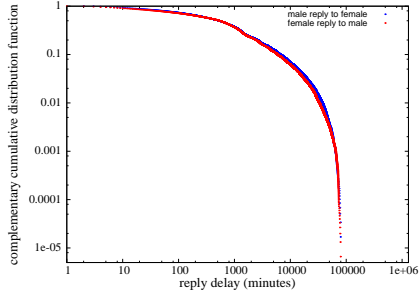}
    \caption{CCDF of the reply delay of messages sent by sample users.}
   \label{fig:replyDelayCCDF}
\end{figure}

We obtained the reply delay distribution for 209,863 messages replied to by users within the dataset and plot it in Figure \ref{fig:replyDelayCCDF}. The reply delay distribution exhibits a log-normal behavior with a cut-off point around 79,424 minutes (approximately 56 days or 8 weeks). Note that the cut-off point is due to the fact that we only have the communication record for each user during the first eight weeks of his/her membership, so the obtained distribution is limited by this factor.  

There is little difference in the reply delay distribution for male and female users. The median reply delays of males and females are 8.9 hours and 9.0 hours, respectively. Most messages were replied to within a short time frame. Around 23.0\% of the messages were replied to within one hour, and 72.6\% of the messages were replied to within 24 hours. On the other hand, there is a small fraction of the messages with a long reply delay of tens of days. For example, about 6.3\% of the messages required a week or more to generate a reply. 

%% file: replyProb2.tex
\section{Message Sending and Replying Behaviors} \label{sec:behavior}

After a user creates an account on the online dating site, he/she can search
for potential dates based on information within the profiles provided by other
users including user location, age, etc. Once a potential date has been discovered, 
the user then sends a message to him/her, which may or may not be replied to by the
recipient. The message sending and replying behaviors of a user are strong indicators 
of what he/she is looking for in a potential partner and reflect the user's actual 
dating preferences.

In this section, we first present the correlation between user send and reply behaviors 
with various user attributes including age, height, income, education 
level, distance, and photo count. We further examine how actual user behavior
deviates from random selection where user attributes (e.g., age, 
height, income, etc) of the recipient of a message are randomly drawn from their 
respective distributions. When appropriate, error bars are provided with a 95\% 
confidence interval.

At the online dating site, a user can provide his/her preferences for potential dates 
in terms of age, location, height, education level, income range, marriage and children 
status, etc. In the design of a recommendation algorithm for potential dates, it is 
important to know whether and to what extent users follow their stated preferences 
in actual dating. The discrepancy between a user's stated preference and his or her
actual dating behavior is often referred to as dissonance in social psychology, and 
has been previously observed \cite{Eastwick08Sex}. In this section, we examine the 
degree of dissonance of online dating in our dataset. In particular, we study to what 
extent users adhere to their stated preferences and how reply probability varies as a 
function of the number of user attributes that match receiver's stated preference.


\subsection{Age}
\begin{figure} 
\subfloat[\label{fig:ageDifferenceSending}]
  {\includegraphics[width=.5\linewidth]{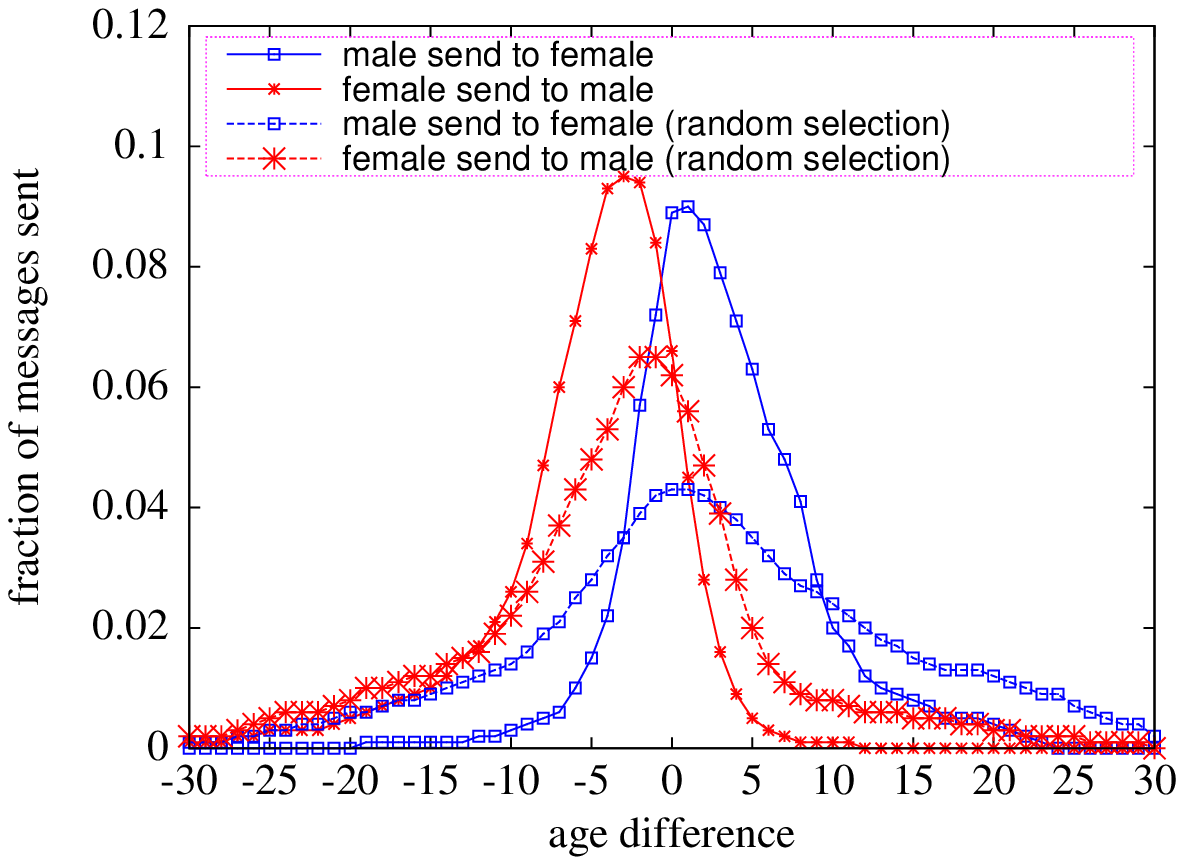}}\hfill
\subfloat[\label{fig:ageDifferenceReply}]
  {\includegraphics[width=.5\linewidth]{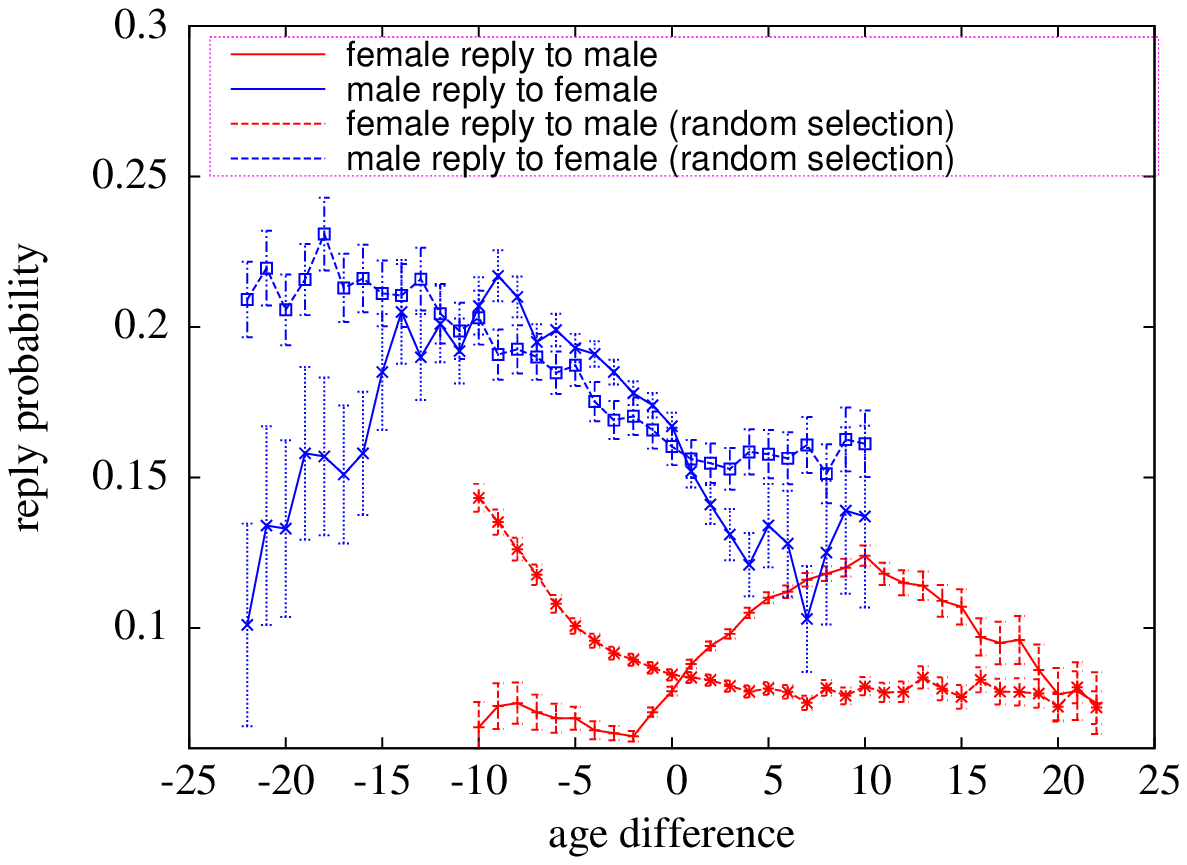}}
\caption{(a) Distribution of age difference between senders and receivers. (b) Reply probability for users with different age difference.}
\label{fig:age}
\end{figure}

Figure \ref{fig:age}\subref{fig:ageDifferenceSending} shows the distribution of the age difference 
between the sender and receiver of all messages sent by the sample users in the dataset. 
The age difference is computed as the sender's age less the receiver's age. While 
the age difference between senders and receives covers a wide range, the preferences 
of males and females are opposite of each other. Males tend to look for younger females 
and the distribution is skewed towards much younger females. On the other hand, females 
tend to look for older males and the distribution is skewed toward older males. The median
age difference is two for messages sent from males to females and -4 vice versa.
Male and female preferences are not random; they look for potential
dates with a smaller age difference than predicted by random selection. 


Figure  \ref{fig:age}\subref{fig:ageDifferenceReply} plots the reply probability as a function of the age 
difference between the sender and receiver of a message. For both males and females, 
the reply probability deviates significantly from the result of random selection,
exhibiting a bell shape mode at a age difference of ten years older and eight years younger,
respectively. Males tend to reply to younger females while females tend to reply to older 
males within a certain range of age difference.

\begin{figure}
\subfloat[\label{fig:male_Send_heat}]
  {\includegraphics[width=.5\linewidth]{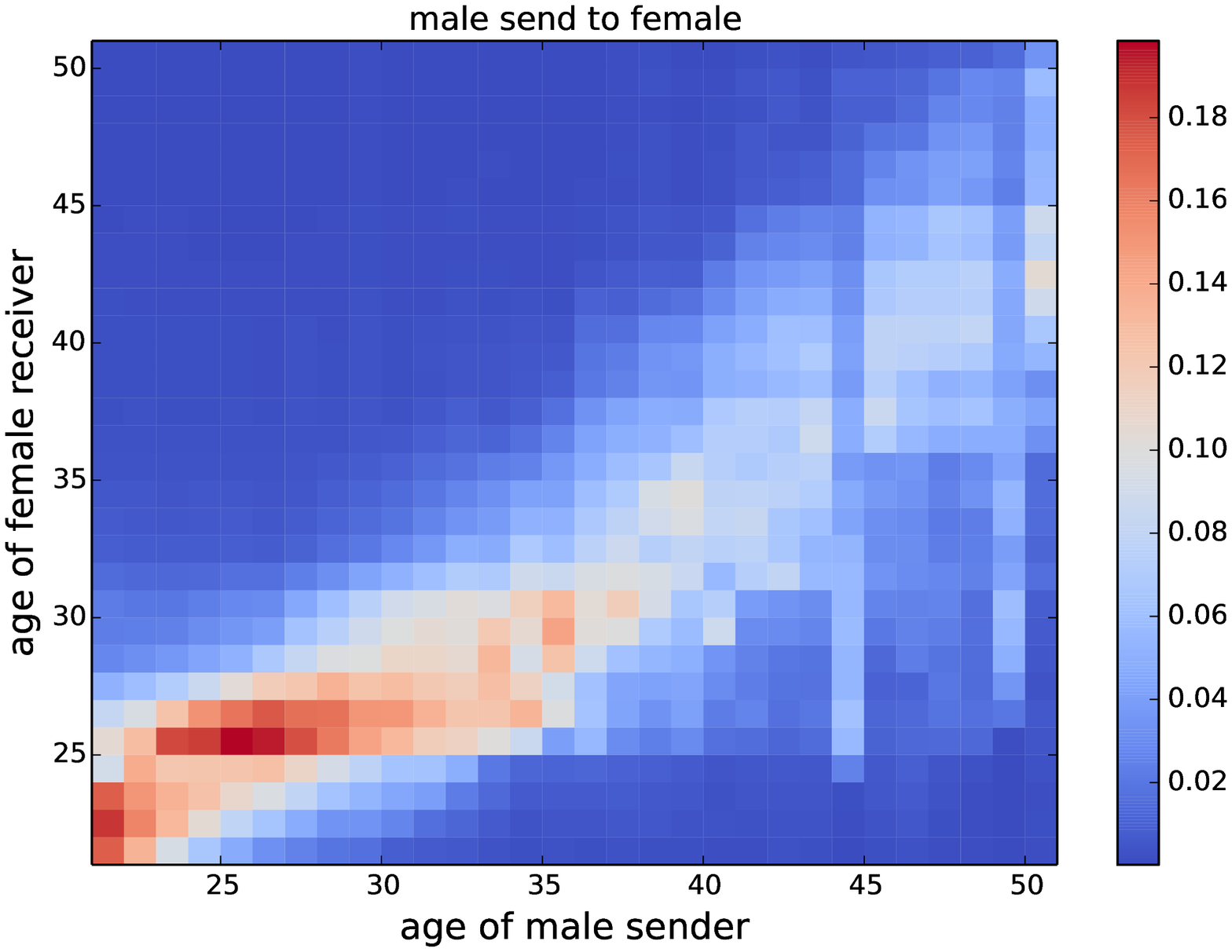}}\hfill
\subfloat[\label{fig:female_Send_heat}]
  {\includegraphics[width=.5\linewidth]{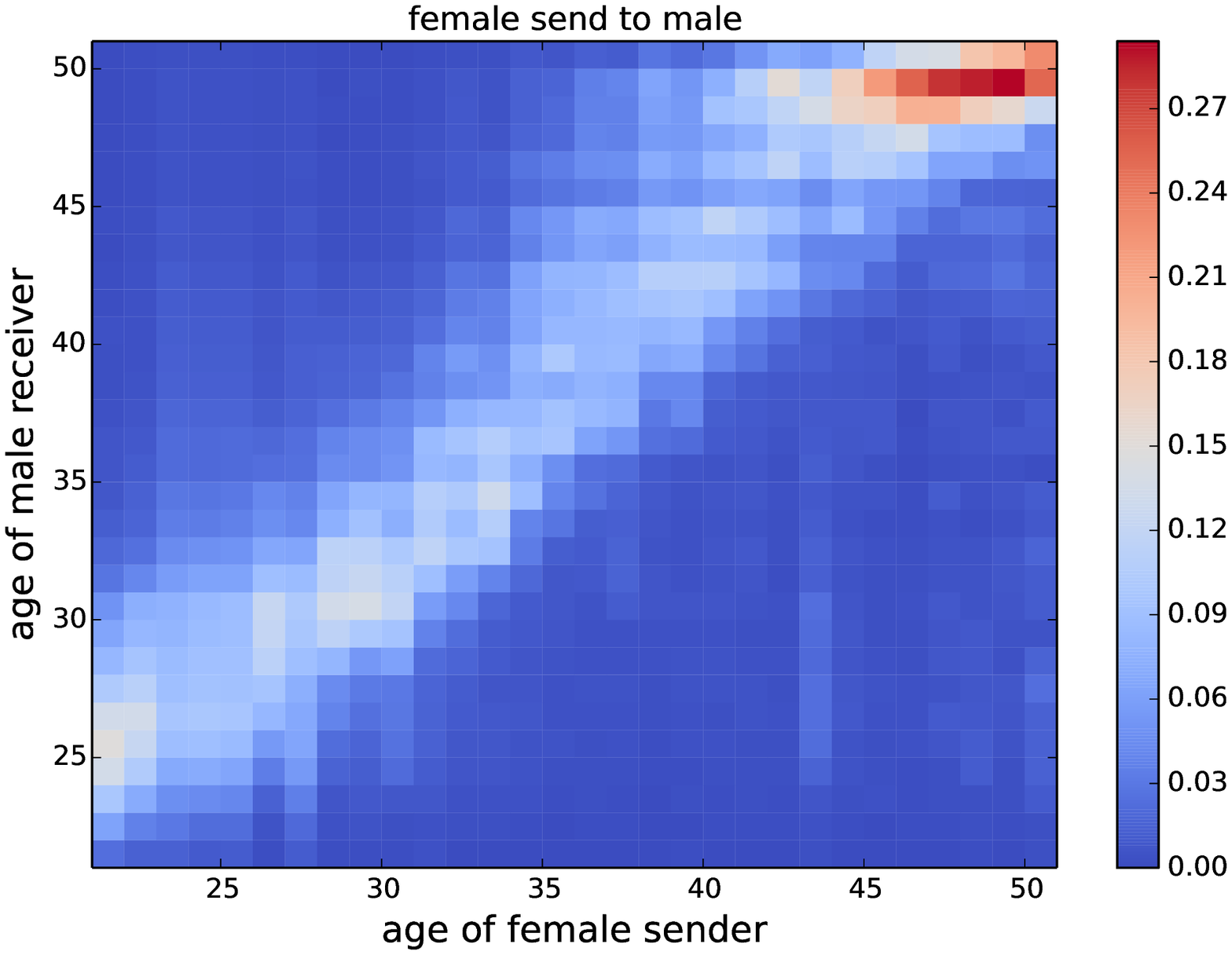}}\hfill
\subfloat[\label{fig:male_Reply_heat}]
  {\includegraphics[width=.5\linewidth]{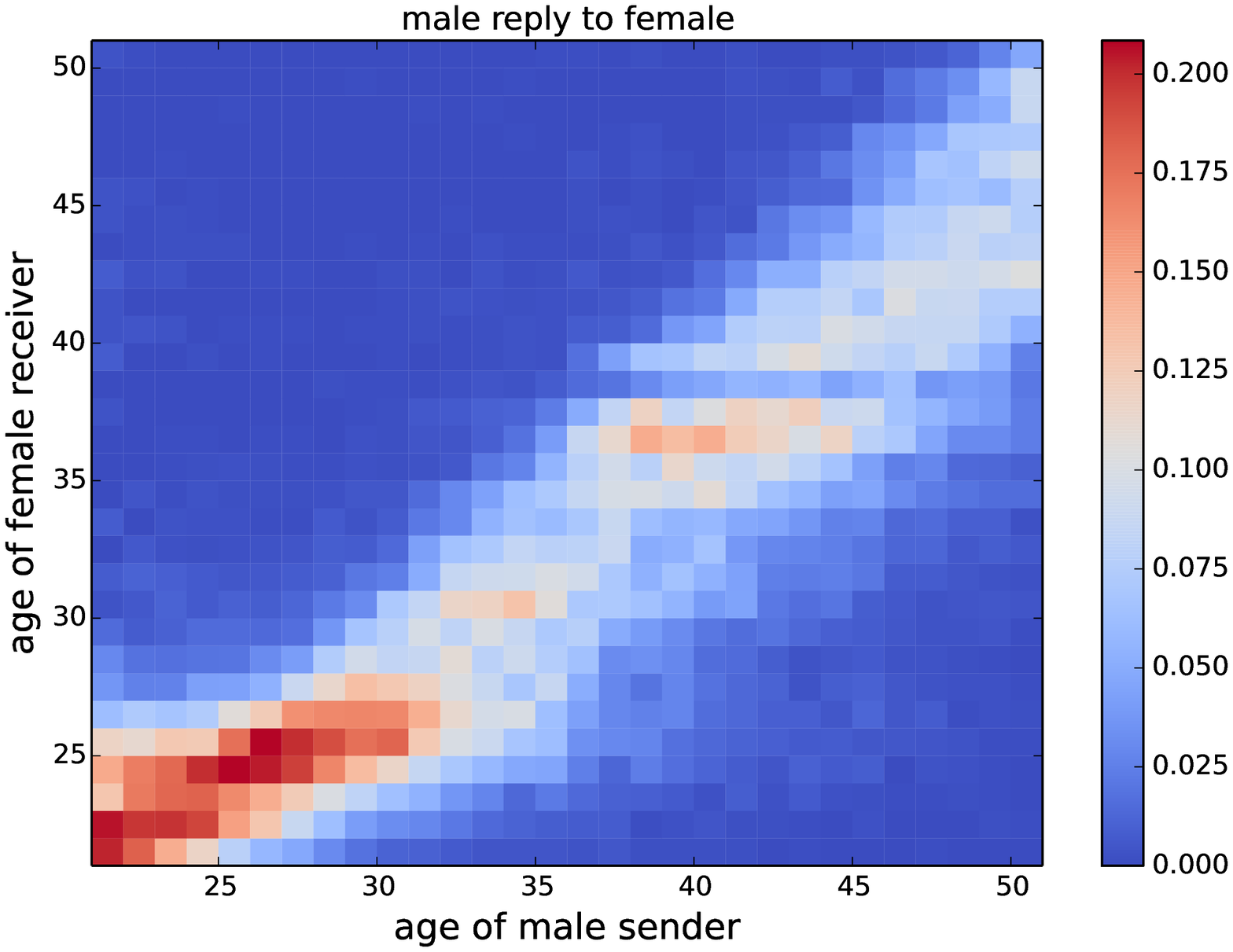}}\hfill
\subfloat[\label{fig:female_Reply_heat}]
  {\includegraphics[width=.5\linewidth]{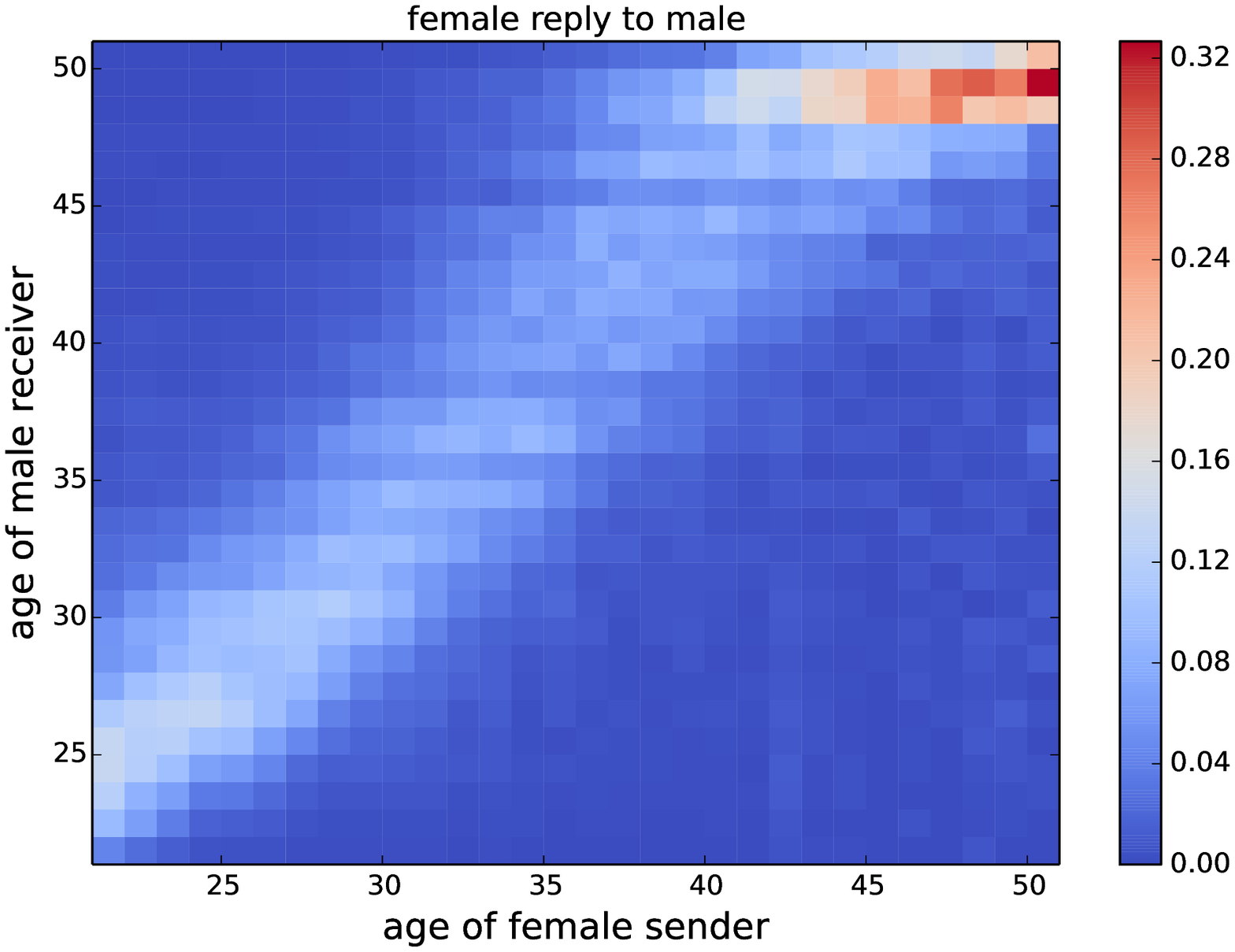}}
\caption{Heat map of fraction of messages and reply probabilities between users of different age: 
(a) fraction of messages sent from males to females, (b) fraction of messages sent from females to 
males, (c) reply probability of males to females, (d) reply probability from females to males.}
\label{fig:Heat_Mapt}
\end{figure}


Figure \ref{fig:Heat_Mapt} depicts the heatmap of the fraction of messages and reply probabilities 
between users of different age. As a male gets older, he searches for and replies to relatively 
younger females. A female in her 20's is more likely to communicate with older males, 
but as a female gets older, she becomes more open towards younger males. This is the cause for the reply 
probability increase in the age difference range from -3 to -10, as shown in 
Figure \ref{fig:age}\subref{fig:ageDifferenceReply}. 
These results are consistent with observations made in \cite{Fiore10Who}.


\subsection{Height}
\begin{figure}
\subfloat[\label{fig:sampleSendHeightDifference}]
  {\includegraphics[width=.5\linewidth]{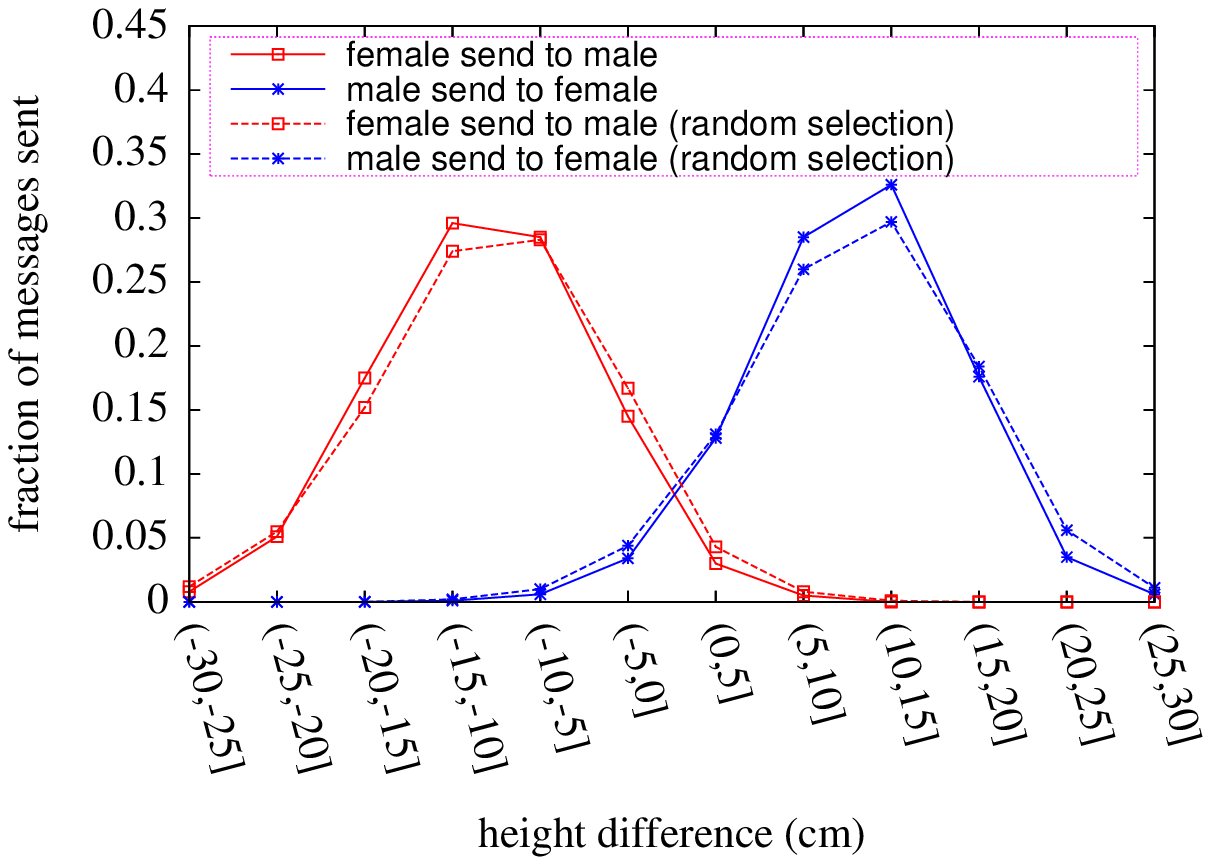}}\hfill
\subfloat[\label{fig:heightDifferenceReplyErrorBar}]
  {\includegraphics[width=.5\linewidth]{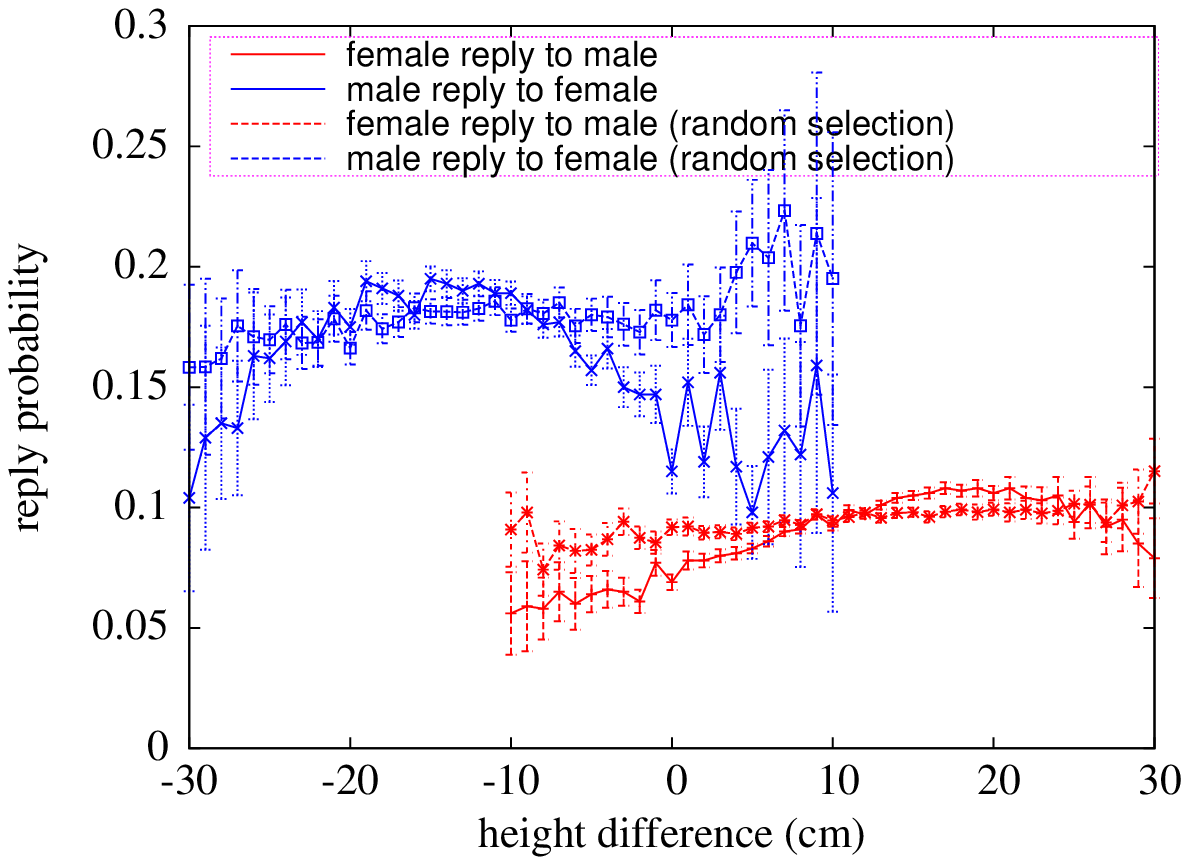}}
\caption{(a) Distribution of height difference between senders and receivers. (b) Reply probability for users with different height difference.}
\label{fig:height}
\end{figure}

Figure \ref{fig:height}\subref{fig:sampleSendHeightDifference} shows the distribution of height difference 
between the sender and receiver of all messages sent by sample users. The height difference 
is computed as the sender's height less the receiver's height. We observe that users' 
message sending behaviors with respect to height closely match those resulting from random 
selection. While it appears that a male tends to look for females shorter than him and a female 
tends to look for males taller than her, this is likely to be a result of random selection rather 
than users' preference.


Figure \ref{fig:height}\subref{fig:heightDifferenceReplyErrorBar} plots the message reply probability 
as a function of height difference between the senders and receivers.  Similarly,
user message reply behavior with respect to height closely match that of 
random selection, and are thus likely to be the result of random selection rather 
than user preference.


%

\subsection{Income}

\begin{figure}
\subfloat[\label{fig:incomeDifference}]
  {\includegraphics[width=.5\linewidth]{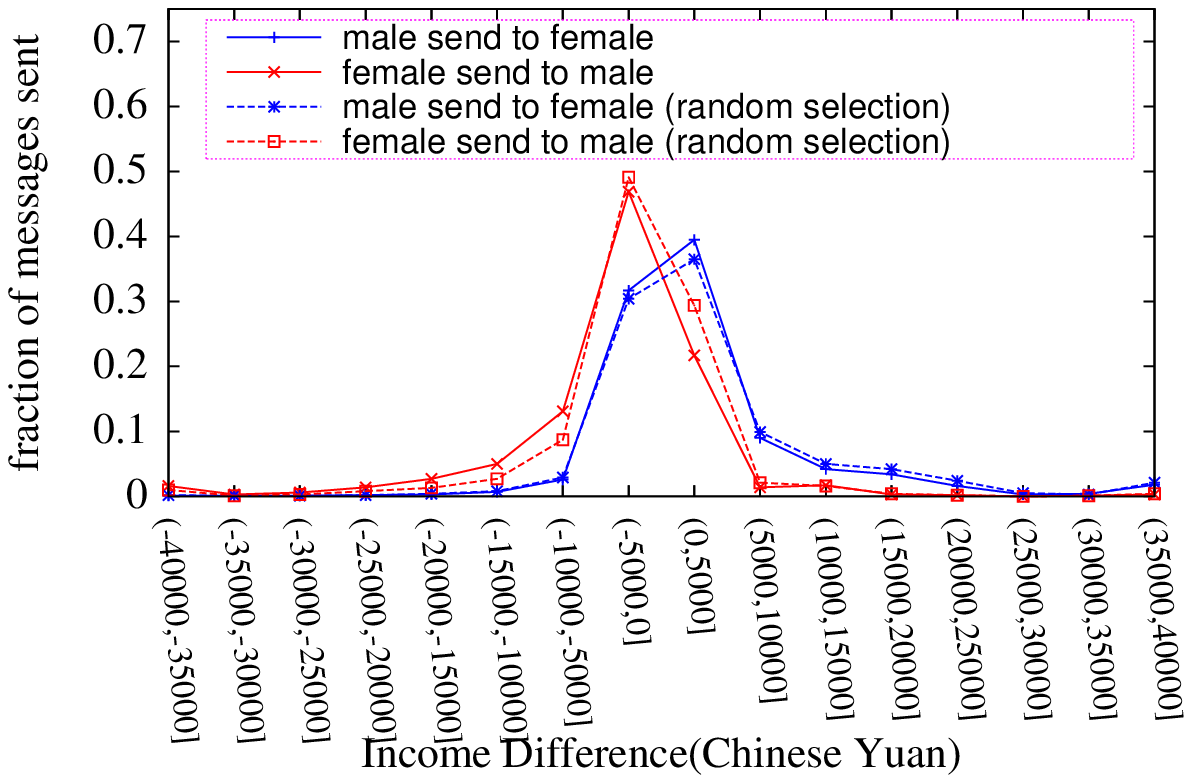}}\hfill
\subfloat[\label{fig:incomeReplyPDF}]
  {\includegraphics[width=.5\linewidth]{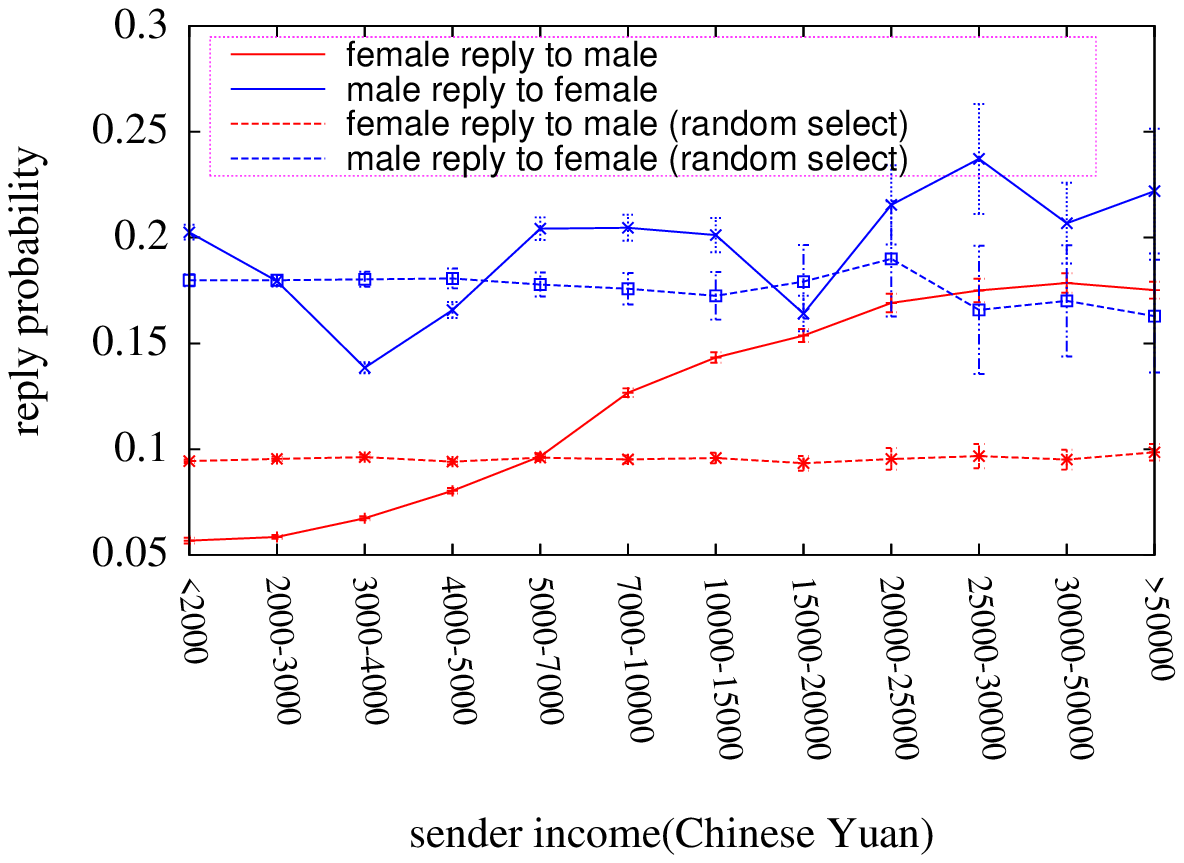}}
\caption{(a) Distribution of income difference between senders and receivers. (b) Reply probability for senders with different incomes.}
\label{fig:income}
\end{figure}

Figure \ref{fig:income}\subref{fig:incomeDifference} shows the distribution of income difference between 
senders and receivers.  A user reports monthly income within a range such as below 2,000, 2,000-3,000 
(all in Chinese Yuan), etc. We take the median value of the 
reported income range as a user's income and the income difference between the sender and 
receiver of a message is computed as the difference sender income and receiver income. 
We observe that user message sending behavior with respect to income closely matches 
that resulting from random selection. While it appears that males tend to send messages to 
females with lower income and females tend to send messages to males with higher income,
this is likely to be a result of random selection and the fact that male incomes are larger
than female incomes rather than users' preference.


Figure \ref{fig:income}\subref{fig:incomeReplyPDF} shows how reply probability varies with 
sender income. The reply probability of female recipients increases with male sender income, deviating 
significantly from the flat line of random selection. There is a strong correlation 
coefficient of 0.90 between the reply probability and male sender income. 
On the other hand, the income of a female does not have as significant an 
effect on the likelihood of her messages being replied to. The reply probability fluctuates 
around the line of random selection. The correlation between the reply probability and 
female sender income is much weaker with a correlation coefficient of 0.50.



\subsection{Education level}

\begin{figure}
\subfloat[\label{fig:educationSendReceiver}]
  {\includegraphics[width=.5\linewidth]{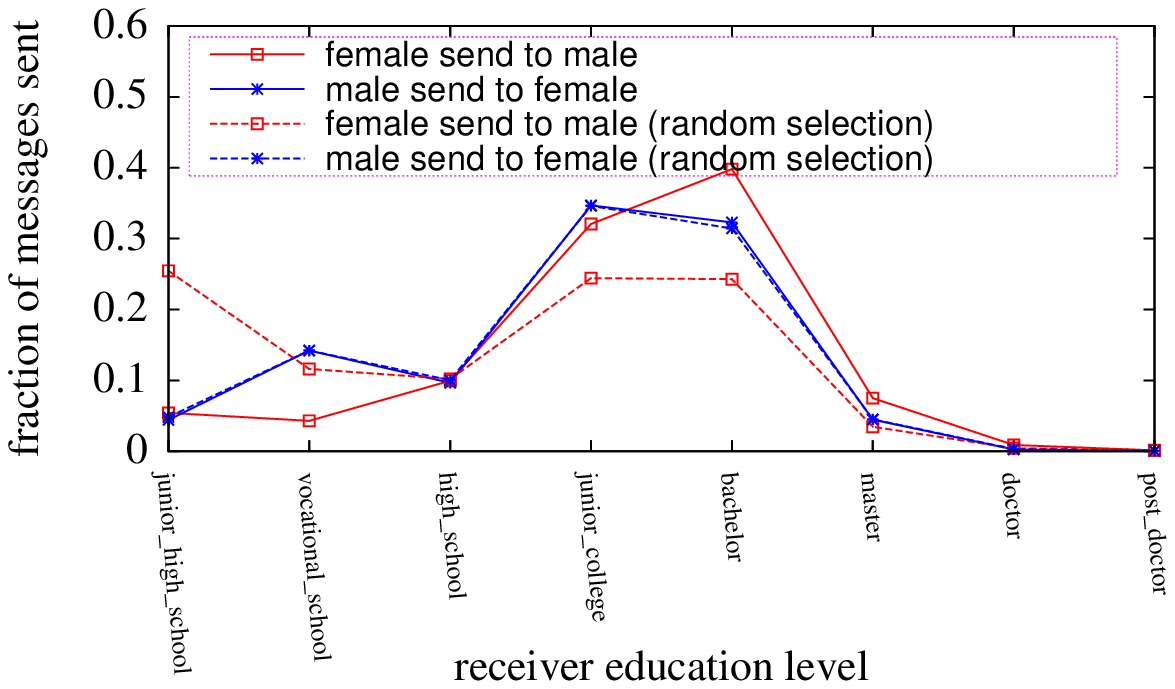}}\hfill
\subfloat[\label{fig:educationReplyErrorBar}]
  {\includegraphics[width=.5\linewidth]{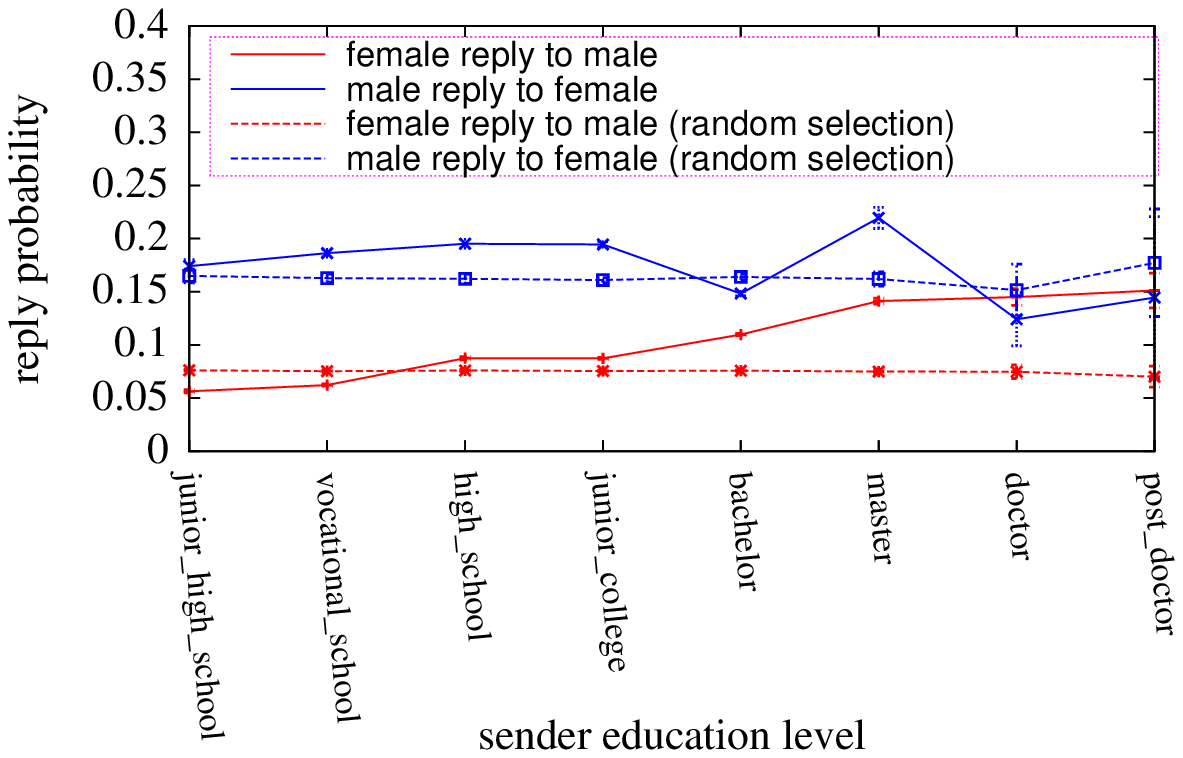}}
\caption{(a) Fraction of messages sent to users of different education levels. (b) Reply probability for messages from users of different education levels.}
\label{fig:education}
\end{figure}

Figure \ref{fig:education}\subref{fig:educationSendReceiver} shows the fractions of messages sent to users of different
education levels. We observe that male behavior closely matches that of random selection, while
female behavior deviates considerably from that of random selection towards higher education 
levels.

Figure \ref{fig:education}\subref{fig:educationReplyErrorBar} shows how reply probabilities vary with 
sender education levels for males and females. The higher the education level of a male sender, 
the more likely his messages will be replied to. The reply probability of a female user deviates
significantly from a random selection.  On the other hand, the education 
level of a female does not have as significant an effect on the likelihood of her messages being 
replied to. The reply probability of male users stays relatively flat across different eduction
levels, similar to that resulting from random selection. 
\\

%

\subsection{Geographic distance}
The geographic distance between two users plays an important role in their online dating behavior. 
As mentioned in Section \ref{sec:dataset}, a considerable portion (46.5\%) of 
the communications occurred between users within the same city. For 
communications between users in different cities, we further study how 
message sending behavior and reply probability varies with the distance between users 
(computed as the straight line distance between the two cities).

\begin{figure}
\subfloat[\label{fig:distanceMessagePDF}]
  {\includegraphics[width=.5\linewidth]{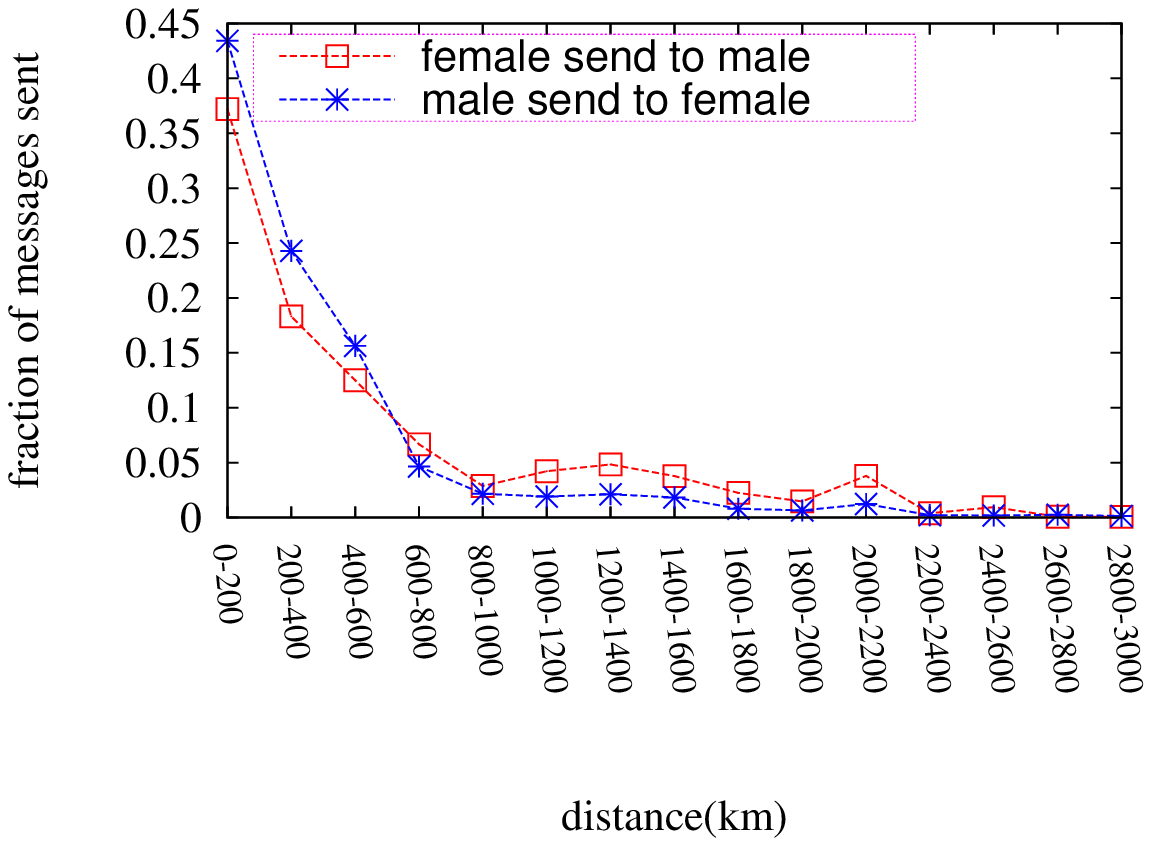}}\hfill
\subfloat[\label{fig:distanceReply}]
  {\includegraphics[width=.5\linewidth]{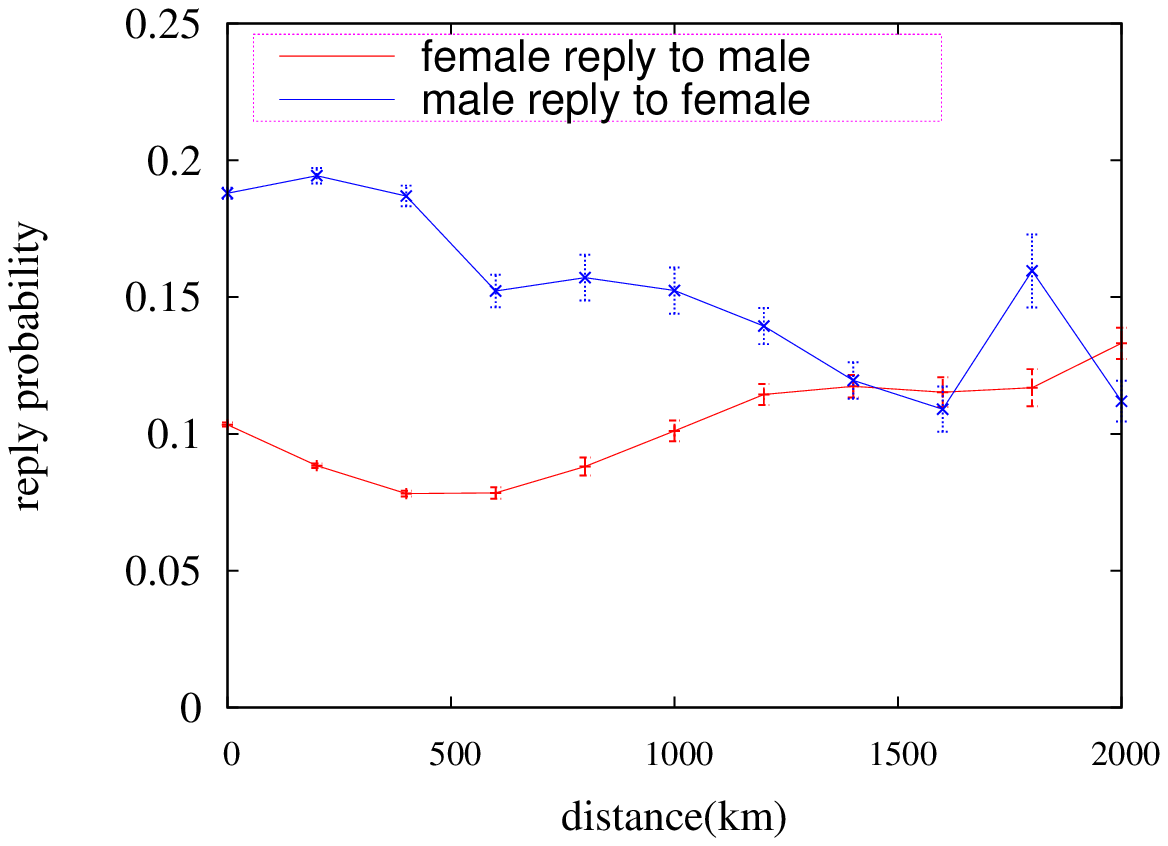}}
\caption{(a) Distribution of messages of different send-receiver distances. (b) Reply probability for users with different distances.}
\label{fig:distance}
\end{figure}


As shown in Figure \ref{fig:distance}\subref{fig:distanceMessagePDF}, in general the fraction of 
messages decreases as the distance between users increases. The messages 
between users of at least 1,000 km apart constitutes only a small fraction (11.7\%) 
of the total number of messages. Note that there is a small increase in the 
fraction of messages between distance 800 and 1,400km for female senders. 

Figure \ref{fig:distance}\subref{fig:distanceReply} depicts how reply probability varies with 
distance between a sender and receiver.
When a male receives a message from a female, the reply probability generally 
decreases with distance between them. For females, the reply probability first 
decreases with distance but increases in the range from 800 to 1,400km.

The increase of the initial message ratio and reply probability of females for the
distance range from 800 to 1,400km is due to the following. There is 
an increasing number of big cities (Shanghai, Beijing, Hong Kong, Chongqing, Guangzhou, 
Xi'an, etc) between many of which the distance falls into this range, and unlike 
males, females are more likely to send and reply to messages between these cities. 

\begin{figure}
\centering
\subfloat[\label{fig:photoCountPDF}]
  {\includegraphics[width=.33\linewidth]{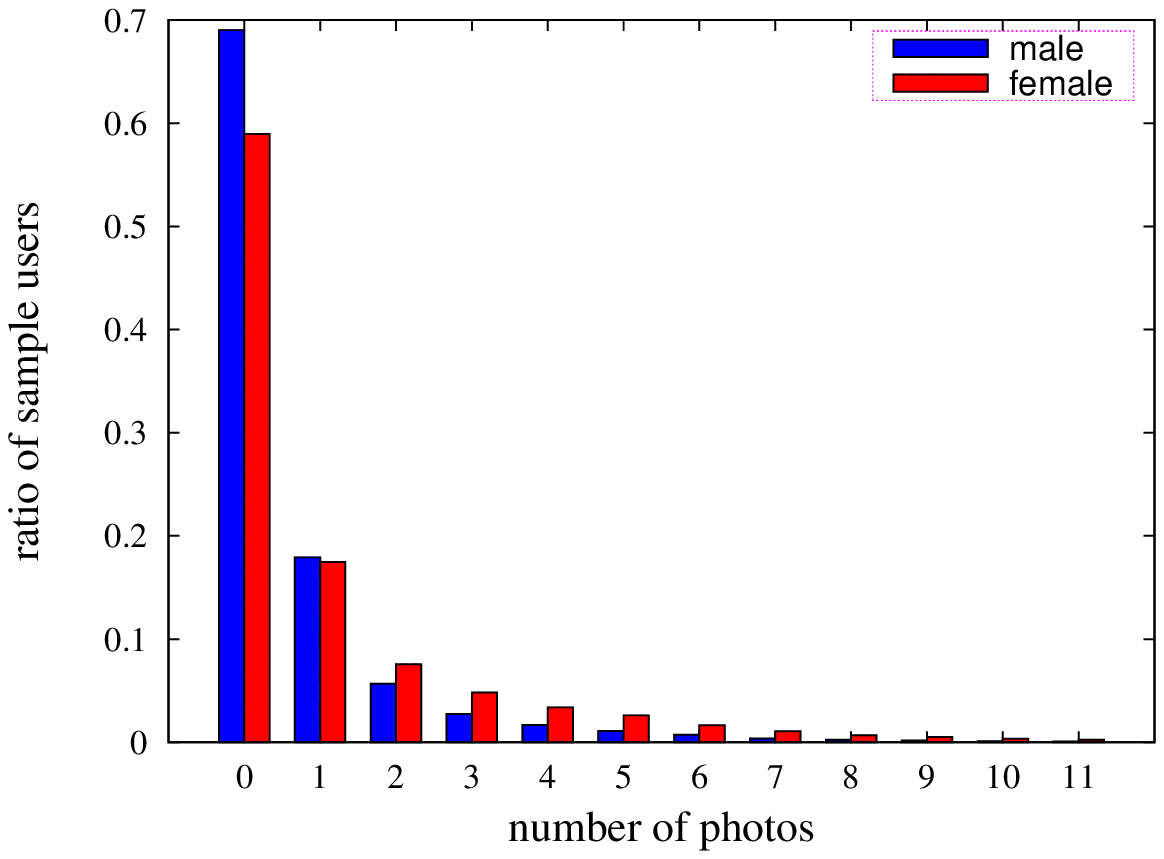}}\hfill
\subfloat[\label{fig:photoReceiveAverage}]
  {\includegraphics[width=.33\linewidth]{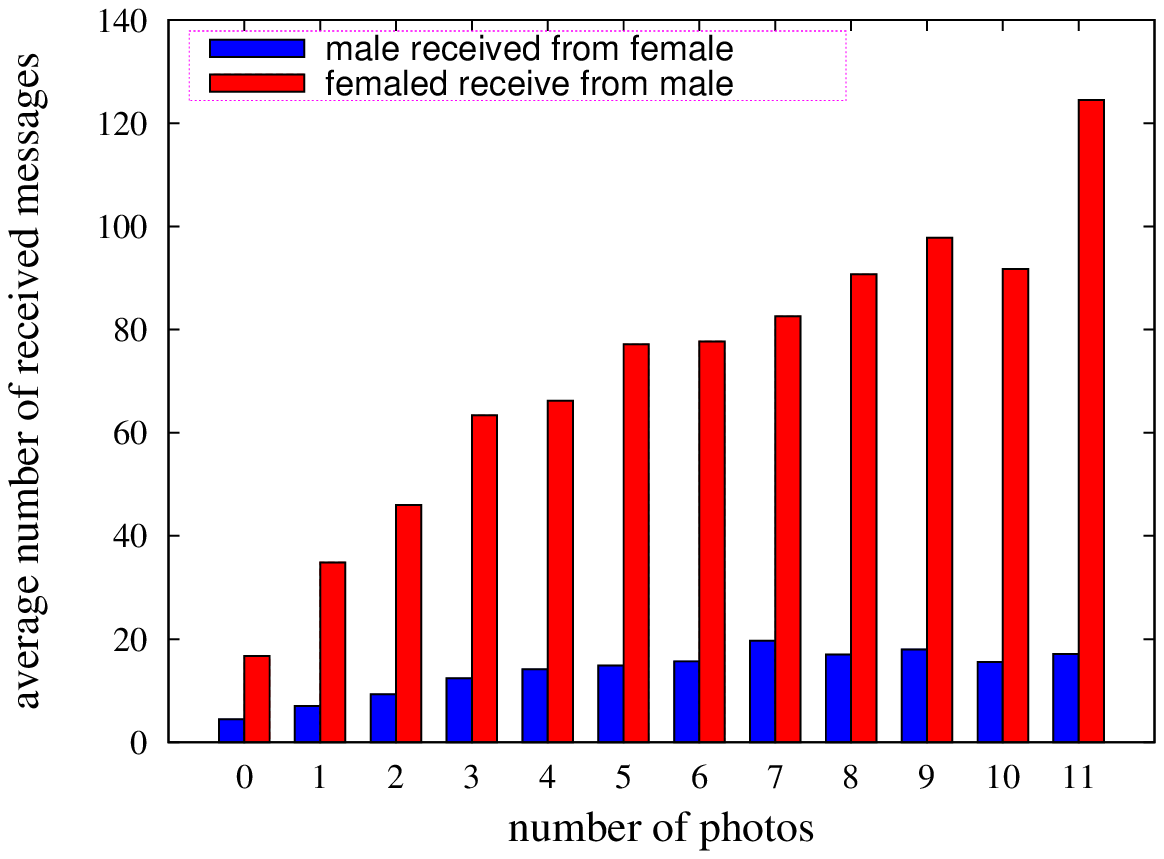}}\hfill
\subfloat[\label{fig:photoReply}]
  {\includegraphics[width=.33\linewidth]{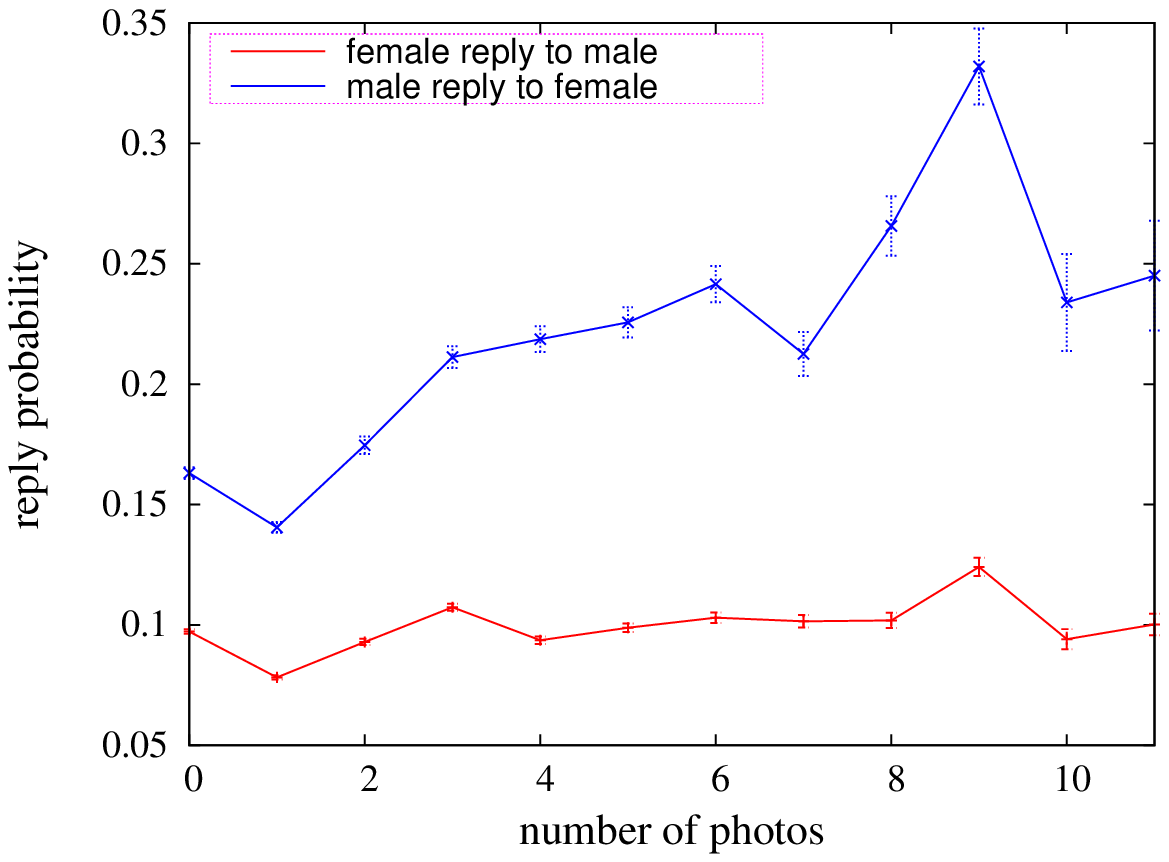}}
\caption{(a)Distribution of users' photo count. (b) Average number of received messages during first eight weeks of their memberships for users with different photo count. (c) Reply probability for users with different photo counts.}
\label{fig:photo}
\end{figure}


\subsection{Photo count}

On the dating site, a user can post photos on his/her profile page. Figure 
\ref{fig:photo}\subref{fig:photoCountPDF} plots the distribution of the number of photos posted 
by a user. A large fraction of users did not post or posted only a small number 
of photos. In our dataset, about 69\% of male users and 59\% of female users 
did not post any photos. Female users tend to post more photos than male users.


As shown in Figure \ref{fig:photo}\subref{fig:photoReceiveAverage}, a user tends to receive more
messages if he/she has posted more photos online, with the trend being 
more pronounced for females than for males. The number of received messages 
by a male user starts to level off after some point.


Figure \ref{fig:photo}\subref{fig:photoReply} shows how message reply probability varies 
with the number of photos posted by the sender. We observe that 
male reply probability tends to increase with the 
number of photos posted by the female sender. Interestingly, when a female 
receives a message, the reply probability remains relatively stable 
as the number of photos of the male sender increases.


\begin{figure}[htb]
\centering
\includegraphics[width=2.5in]{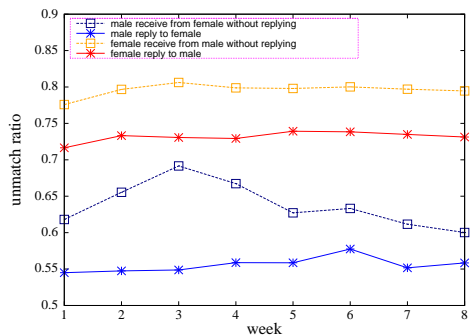}
    \caption{Fraction of reply messages that violate senders' stated preference as a function of time.}
   \label{fig:sampleReplyUnMatchWeekly}
\end{figure}

\subsection{Stated preference versus actual behavior}

On the online dating site in our study, a user can specify a set of attributes that he/she 
is looking for in a date, including age range, geographic location, height range, marriage
status (never married, divorced), education level, income range, house ownership, and
children status (no children, children living with user, children not living with user). 

There is often considerable discrepancy between a user's stated preference and his or her 
actual dating behavior \cite{Eastwick08Sex}. 
Therefore, it is important to understand users' true dating preferences in order to make better 
dating recommendations. In this section we study to what extent users adhere to their stated 
preferences and how reply probability varies as a function of the number of user attributes that 
match receiver's stated preference.

\begin{figure}
\subfloat[\label{fig:maleReplyAttributesUnMatch}]
  {\includegraphics[width=.5\linewidth]{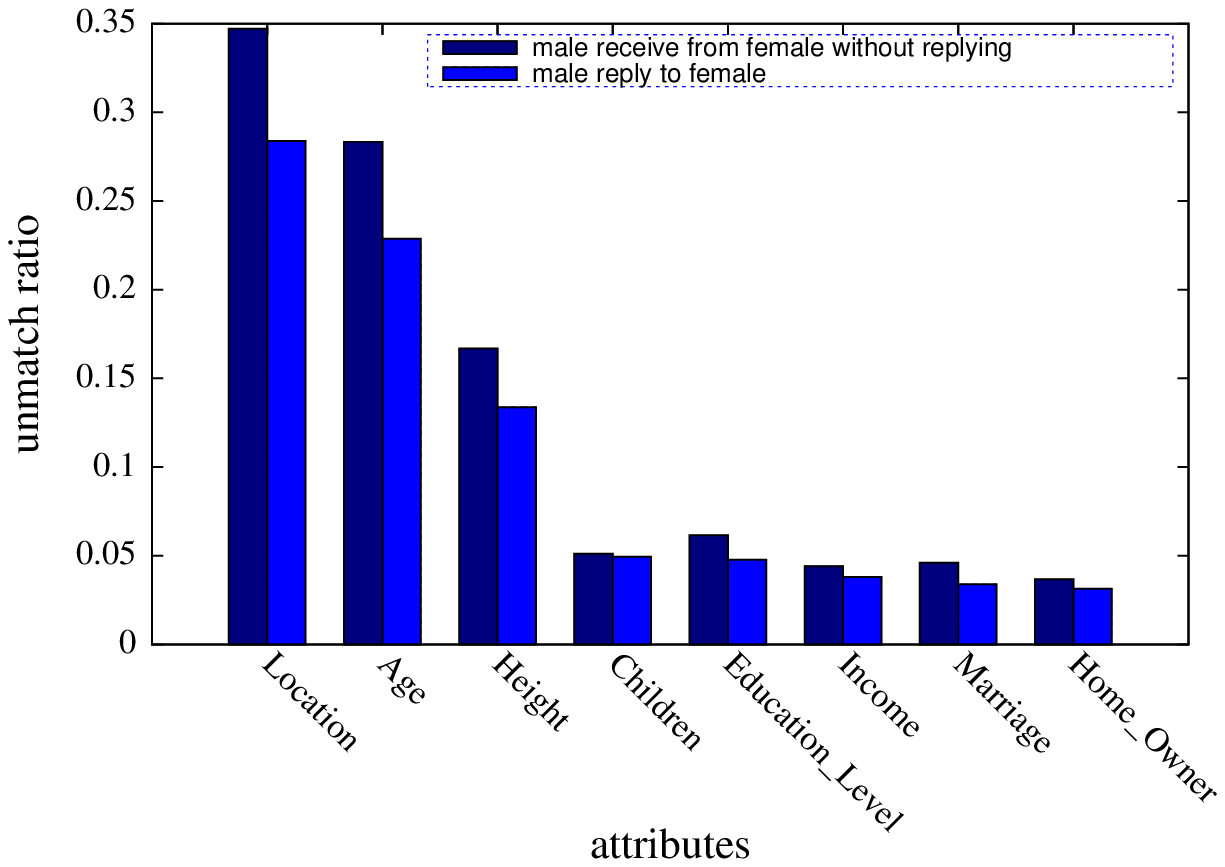}}\hfill
\subfloat[\label{fig:femaleReplyAttributesUnMatch}]
  {\includegraphics[width=.5\linewidth]{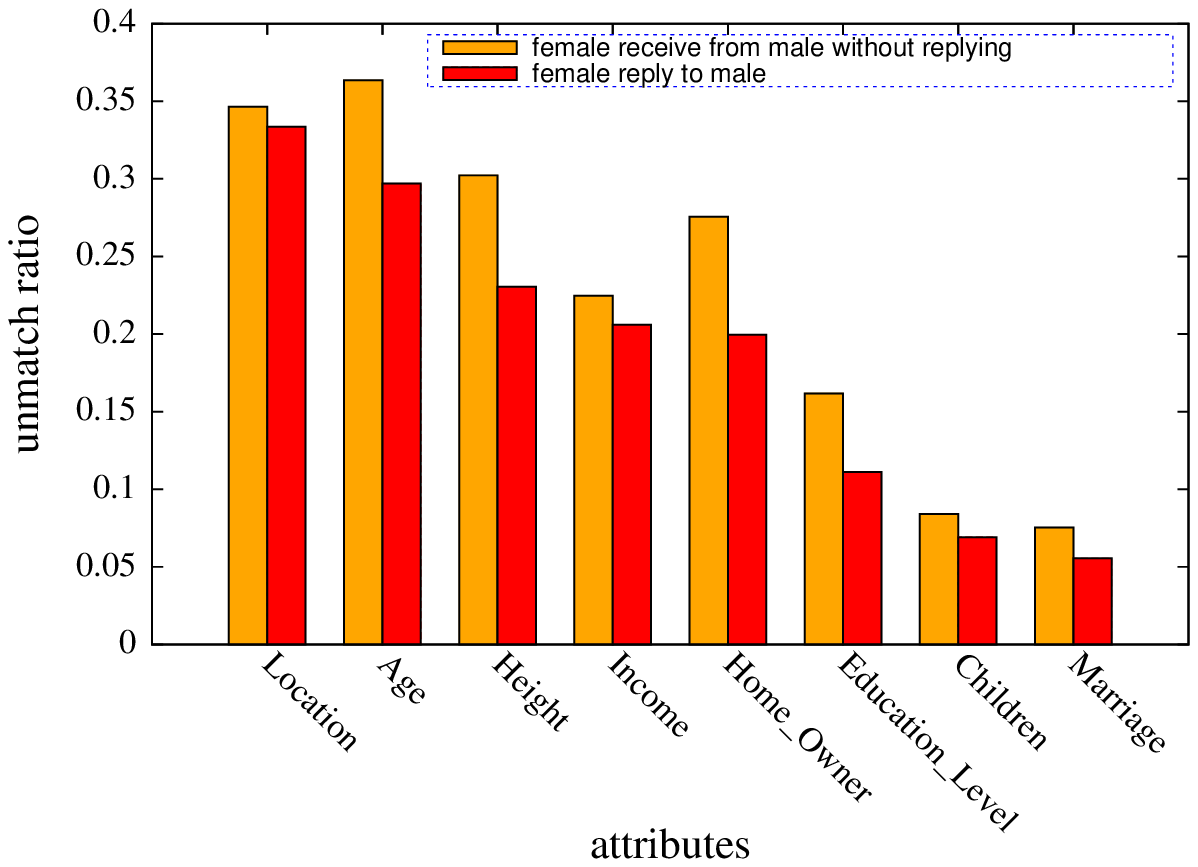}}
\caption{Unmatch ratio of different user attributes in reply messages sent by (a) male users and (b) female users.}
\label{fig:replyUnMatchAttributes}
\end{figure}

Figure \ref{fig:sampleReplyUnMatchWeekly} shows the fraction of replied and unreplied messages whose 
senders do not satisfy the recipients' stated preference in at least one user attribute as a function 
of time. We refer to this fraction as the unmatch ratio. Among all replied messages, the unmatch ratio 
is around 55\% for males and more than 70\% for females. The discrepancy between a user's stated dating 
preference and his/her actual behavior is prevalent, with female users showing more flexibility than male 
users. Actually, during the eight week period since their memberships, only 17.0\% of male users and 6.6\% 
of female users had strictly followed their stated preferences when replying to a sender. We also observe 
that the unmatch ratio is larger for messages not replied to than those replied to. This indicates 
that out of the population of users that send messages, replies are more likely to go to those whose 
attributes come closest to the preferences of the receivers.

Figures \ref{fig:replyUnMatchAttributes}\subref{fig:maleReplyAttributesUnMatch} and \ref{fig:replyUnMatchAttributes}\subref{fig:femaleReplyAttributesUnMatch}  show the unmatch 
ratio for each user attribute in a decreasing order for both male and female users, respectively. We observe
that males and females share the same top-three most violated user attributes: age, location and height. 
For male users, the unmatch ratios of other attributes are all very low (below 5\%), while female users are 
most strict with marriage and children status, as well as education level of the male senders. For each 
attribute, the unmatch ratio is larger for messages not replied to than those replied to, indicating 
that replies are more likely to go to users whose attributes come closest to the preferences of the receivers.

\begin{figure} 
\centering
\includegraphics[width=3in]{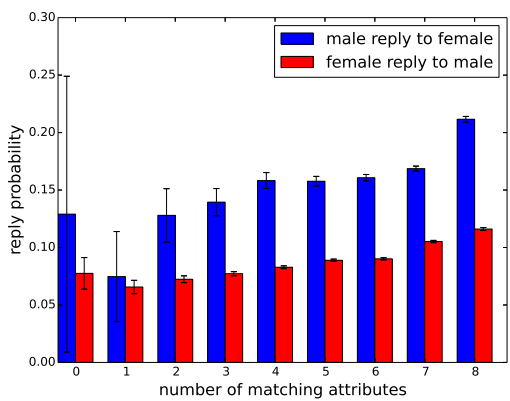}
\caption{Reply probability as a function of the number of user attributes that matches receiver's 
stated preference.}
\label{fig:Rep_Num}
\end{figure}



Figure \ref{fig:Rep_Num} shows how male and female reply probabilities vary as a function 
of the number of sender attributes that match the receiver's stated preference. The margin 
of error is provided with a 95\% confidence level. We observe that except for the case where
is no matching attribute, the reply probability increases with the number of matched user attributes, 
indicating that both males and females tend to reply to senders whose attributes best match 
their stated preferences. Note that although the reply probability for zero matching user attribute 
is larger than that for one matching user attribute, the sample size of users with zero matching 
attribute is rather small and thus the corresponding margin of error is too large to make the 
calculation statistically sound.

\begin{figure}
\subfloat[\label{fig:male_Rep_feature}]
  {\includegraphics[width=.5\linewidth]{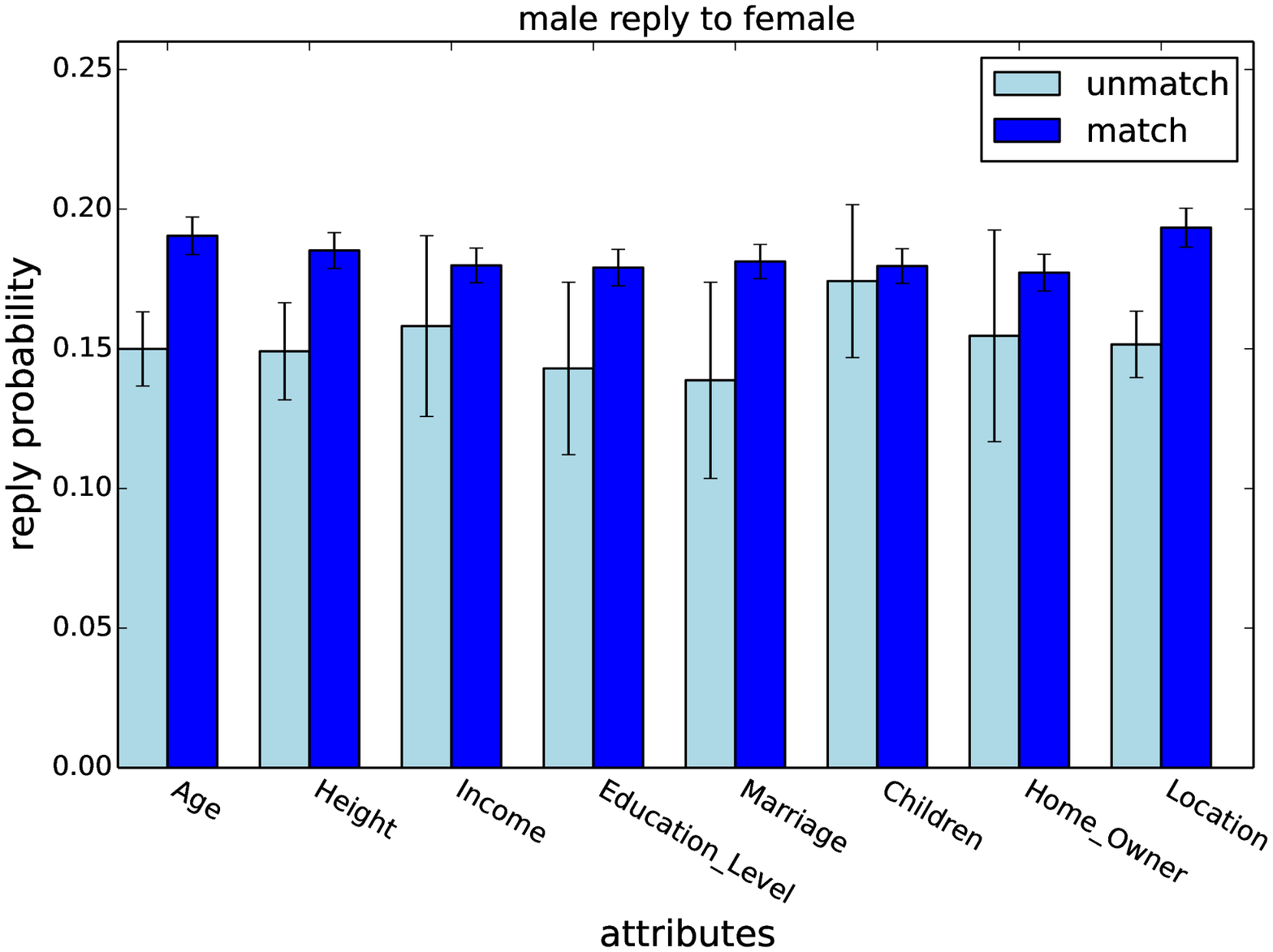}}\hfill
\subfloat[\label{fig:female_Rep_feature}]
  {\includegraphics[width=.5\linewidth]{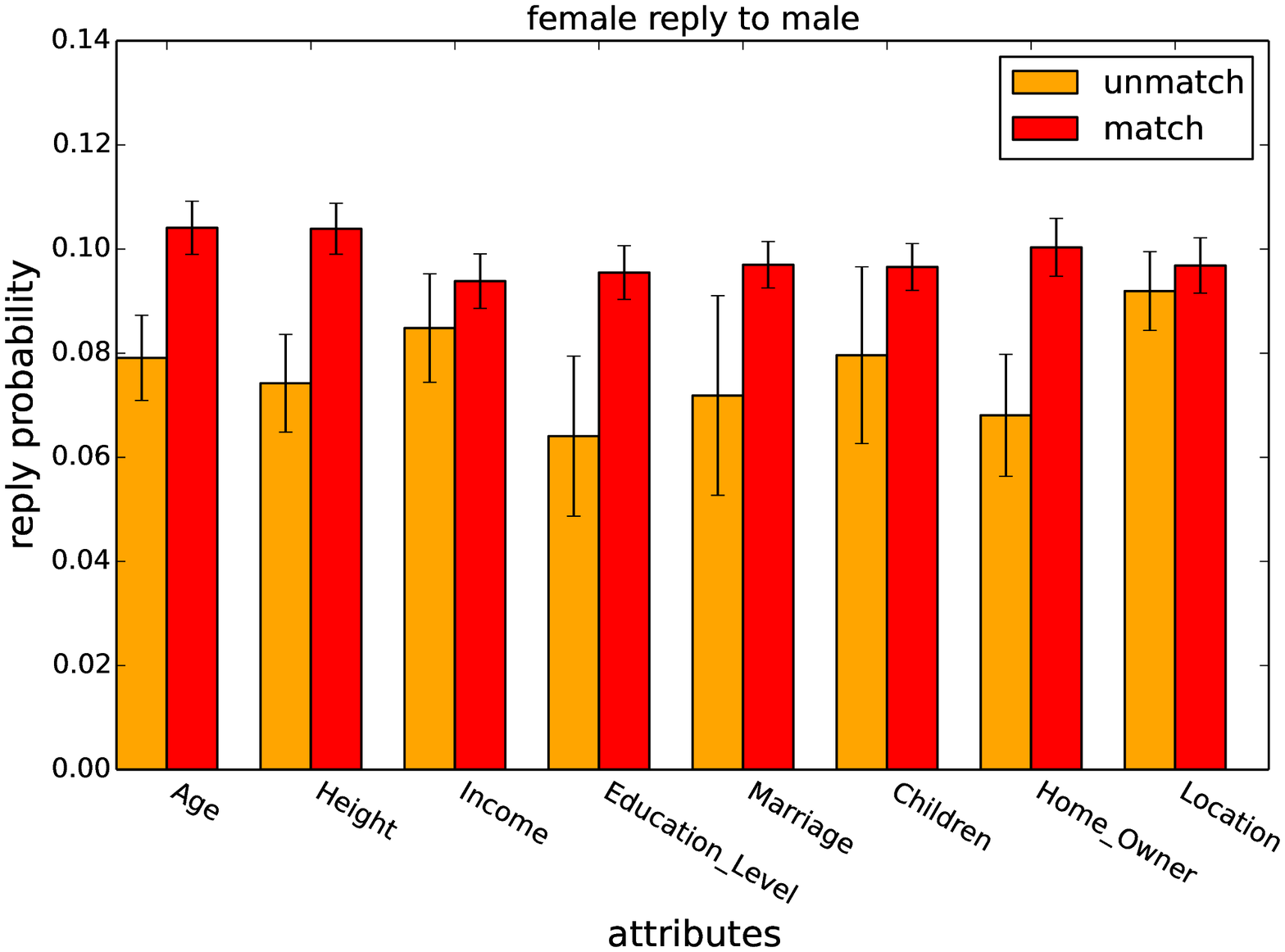}}
\caption{Reply probabilities for scenarios where sender's attribute matches or does not match the receiver's stated preference.}
\label{fig:reply_Feature}
\end{figure}

Figure \ref{fig:reply_Feature}\subref{fig:male_Rep_feature} and Figure \ref{fig:reply_Feature}\subref{fig:female_Rep_feature} compare the reply probabilities
of two different scenarios where a sender's attribute matches or does not match the receiver's stated 
preference, respectively. As expected, we observe that for both males and females, the reply probability 
is larger when the sender's attribute matches the receiver's stated preference.

\section{Discussion} \label{sec:discussions}

Part of our results on user messaging behavior align with notions in social and evolutionary 
psychology \cite{Buss89Sex} \cite{Eagly99Origin} \cite{Luo05Assortative}. Males tend to look for younger 
females but do not seem to care much about the socioeconomic status such as income and 
education level of a potential date. On the other hand, females tend to look for older males 
and place more emphasis on the socioeconomic status of a potential date. Moreover, we observe that
as a male gets older, he searches for relatively younger females. A female in her 
20's is more likely to look for older males, but as a female gets older, she becomes more open 
towards younger males.

Online dating sites significantly increase the level of access to potential dates in terms of
geographic locations from traditional means. In our dataset, a considerable fraction (53.5\%) 
of the initial messages traversed across city boundaries while the remaining 46.5\%  occurred 
between users in the same city. Users still prefer dates in close proximity. For inter-city messages,
the sending volume and reply rate quickly decrease as users live farther apart. Compared to 
male users, females are more likely to send and reply to messages between distant big cities 
(e.g., Beijing, Shanghai, Hong Kong, Guangzhou, Xi'an, etc). 

On the online dating site, a user can post his/her own photos and view other users' photos. 
But profile photos affect male and female's messaging behaviors differently. Females with a larger number 
of photos are more likely to invite messages and secure replies from males, but the photo count of males 
does not have as significant effect in attracting contacts and replies. 

In the analysis of users' dating preferences, our results show that it is important to differentiate 
between user dating preferences and the results of random selection. Some user behaviors in 
choosing attributes in a potential date may largely explained by random selection. For example, while 
it appears that a male tends to look for females shorter than him and a female tends to look for males 
taller than her, the message sending and replying behaviors of both genders closely approximate those 
resulting from random selections, showing that these may be partly due to random selection rather than 
users' true preferences. Similar observations have been made for the behaviors of male users in 
terms of choosing the income and education level of a potential date, while the corresponding female 
behaviors deviate significantly from the random selection and thus reflect their true preferences.

Our results also show that there is significant level of discrepancy between a user's stated 
dating preference and his/her actual online dating behavior. A fairly large fraction of messages
are sent to or replied to users whose attributes do not match the sender or receiver's stated
preferences. Females tend to be more flexible than males in following their stated preferences 
when sending and replying to messages. Both male and female users share the same top-three most 
violated user attributes: age, location and height. For male users, the unmatch ratios of other 
attributes are all very low (below 5\%), while female users are most strict with marriage and children 
status, as well as the education level of the male senders.  For both males and females, out of the 
population of users that send messages, replies are more likely to go to users whose attributes come 
closest to the preferences of the receivers.

%% file: conclusion.tex
\section{Conclusion} \label{sec:conclusion}
We study how people' online dating behaviors correlate with various user 
attributes using a large real-world dateset from a major online dating site 
in China.  Many of our results align with notions in social and evolutionary 
psychology. In particular, males tend to look for younger females while females 
put more emphasis on the socioeconomic status (e.g., income, education level) 
of a potential date. Moreover, geographic distance between two users and the 
photo count of users play an important and different role in dating behaviors
of males and females. Our results show that it is important to differentiate 
between users' true preferences and the results of random selection. Some user 
behaviors in choosing attributes in a potential date may be a result of random 
selection. Our results also show that there is significant discrepancy between 
a user's stated dating preference and his/her actual online dating behavior. 
Our study provides a firsthand account of the user online dating behaviors in 
China, a country with a large population and unique culture. These results 
on users' dating preference can provide valuable guidelines to the design 
of recommendation engine for potential dates.

\section*{Acknowledgement}
This work was supported by the NSF grant CNS-1065133 and ARL Cooperative
Agreement W911NF-09-2-0053. The views and conclusions contained in this
document are those of the authors and should not be interpreted as
representing the official policies, either expressed or implied of the NSF, ARL, or the U.S. Government.